\documentclass[11pt]{article}

\usepackage[final]{acl}

\usepackage{times}
\usepackage{latexsym}

\usepackage[T1]{fontenc}
\everydisplay{\small}

\usepackage[utf8]{inputenc}
\usepackage{enumitem}
\usepackage{CJKutf8}          

\usepackage{microtype}
\usepackage{inconsolata}
\usepackage{caption}
\usepackage{graphicx}
\usepackage[table]{xcolor}

\usepackage{svg}
\usepackage{amsmath}
\usepackage{algorithm}
\usepackage{algorithmic}
\usepackage{tabularx}
\usepackage{caption} 
\usepackage{multirow}
\usepackage{booktabs}

\title{What Makes an Ideal Quote?\\Recommending “Unexpected yet Rational” Quotations via Novelty}

\author{Bowei Zhang\textsuperscript{1}, Jin Xiao\textsuperscript{1}, Guanglei Yue\textsuperscript{1}\\
\textbf{Qianyu He\textsuperscript{2}, Yanghua Xiao\textsuperscript{2}, Deqing Yang\textsuperscript{1}, Jiaqing Liang\textsuperscript{1}\thanks{Corresponding author}}
  \\ \textsuperscript{1}School of Data Science, Fudan University,\\ 
  \textsuperscript{2}Shanghai Key Laboratory of Data Science, College of Computer Science and Artificial Intelligence\\
  \texttt{\{bwzhang24, jinxiao23, glle24, qyhe21\}}@m.fudan.edu.cn,\\
  \texttt{\{shawyh, yangdeqing, liangjiaqing\}}@fudan.edu.cn  \\
}

\begin{document}
\maketitle

\begin{abstract}
Quotation recommendation enriches writing by suggesting quotations that fit a given context, but prior systems largely focus on topical relevance and overlook what makes quotes memorable.
Based on a user study, we find that preferred quotations are often \emph{unexpected yet rational}, motivating the goal of selecting quotes that are contextually novel while semantically coherent.
We propose \textsc{NovelQR}, which (1) uses a generative label agent to map quotations and contexts into multi-dimensional deep-meaning labels for label-enhanced retrieval, and (2) reranks candidates with a token-level novelty estimator that mitigates auto-regressive continuation bias. Experiments on bilingual datasets across diverse domains show that \textsc{NovelQR} is preferred by human judges and improves overall recommendation quality over strong baselines, while achieving competitive novelty estimation.
(Code: \href{https://github.com/Chang-pw/NoQuote}{Github link})


\end{abstract}
\section{Introduction}

\vspace{0.2em}
\begin{center}
        \textit{``Poetic language must appear strange and wonderful.''}
\end{center}
    \begin{flushright}
------\underline{Aristotle}
\end{flushright}
\vspace{0.2em}


Famous quotations~\cite{Quote} play an important role in academic writing and daily communication, as they can ``provide authority for arguments and enhance persuasiveness'' and ``add color and aesthetics to articles''~\cite{maclaughlin2020contextbasedquotationrecommendation}. An appropriate quotation not only helps readers understand complex ideas more accurately, but also adds aesthetic feeling. Therefore,  recommendation of high-quality quotations has become an important task in natural language generation.

This raises a fundamental question: \textbf{what makes an ideal quote?}
Building on Shklovsky’s \textit{Defamiliarization theory}~\cite{defam}, ``\emph{Art aims to renew perception by making the familiar unfamiliar, slowing down understanding and provoking reflection}.'' In this sense, an ideal quotation should not merely restate a point, but challenge habitual thinking and invite deeper interpretation. Related theories in communication and linguistics, such as \textit{Closure theory}~\cite{closure},  suggest that a writing technique prompting deeper thinking and
enhancing aesthetic appeal is the \emph{unfamiliar, complex}, and
\emph{profound} content.
\begin{figure}[t]
    \centering
    \includegraphics[width=1.0\linewidth]{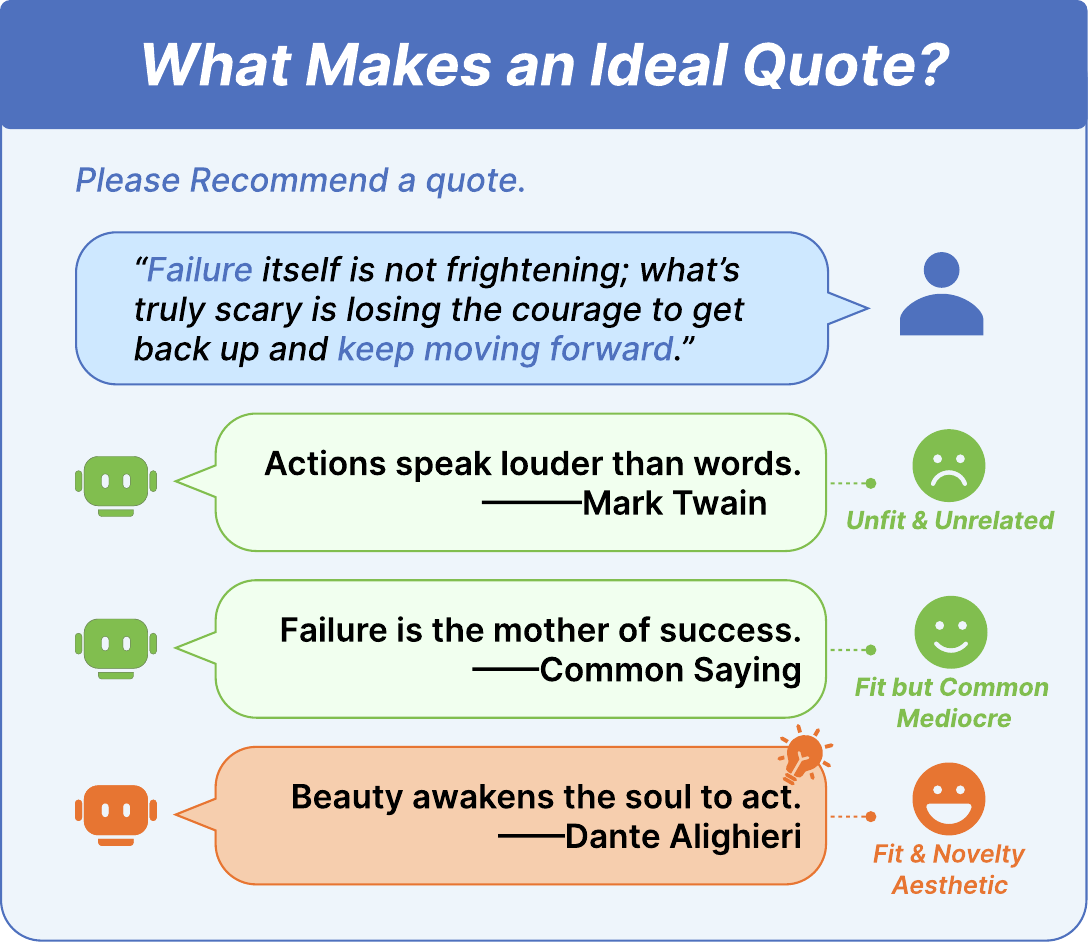}
    \caption{An ideal quote should not only fit the context, but also be novel, adding aesthetic value to writing. As shown in the third example, the best quote often feels unexpected at first, but makes perfect sense in context.}
    \label{fig:intro}
\end{figure}

To examine whether users truly prefer ``unfamiliar'' quotations, we conduct a large-scale user study and controlled behavioral experiments. The results show that, among rationally appropriate options, participants systematically favor more novel quotations and treat novelty as a complementary dimension of quotation quality. We therefore define an ideal quote as ``\textbf{unexpected yet rational}'' (Figure~\ref{fig:intro}): readers may feel briefly puzzled when first encountering the third recommended quote, but then experience a sudden sense of insight once they relate it to the context. Such quotations deepen the expressive power of the context while avoiding clich\'es and mediocrity.

With this defamiliarization- and user-study-–driven view of what constitutes a high-quality quote, we revisit quotation recommendation~\cite{Quote,new1,new2,new3,new4,new5}. Prior systems mostly reduce the task to semantic matching over quote text (e.g., QuoteR~\citep{quoter},  QUILL~\cite{xiao2025quillquotationgenerationenhancement}), emphasizing surface-level rationality while overlooking deeper meanings and the ``unexpected'' dimension. Our analysis shows that even strong LLMs struggle to infer deep meanings from quotations in isolation, and that naive logit-based novelty metrics suffer from an \emph{auto-regressive continuation bias}, such as surprisal~\citep{surp} and KL-divergence~\citep{kl}. These observations motivate a formulation that (1) retrieves quotations in \textbf{a deep semantic space} reflecting their underlying intents, and (2) measures contextual novelty at the \textbf{token level} while mitigating \emph{continuation bias}.

In summary, achieving high-quality quotation recommendation requires addressing two key challenges:
(1) capturing the deep meanings and intents behind quotations, and
(2) measuring novelty while mitigating \emph{continuation bias}.

To address these challenges, we propose \textsc{NovelQR}, a novelty-driven, retrieval-augmented framework for quotation recommendation. A label enhancement module first builds a deep-meaning quotation knowledge base using a generative label agent that interprets each quote into multi-dimensional labels. These labels are used to derive deep-meaning embeddings and support fine-grained hard filtering to ensure semantic rationality. Given a user context, we retrieve candidate quotations by deep-meaning similarity, then apply a token-level novelty estimator that focuses on ``novelty tokens'' to mitigate \emph{continuation bias}. Finally, we integrate novelty, popularity, and matching signals into a unified scoring function to re-rank candidates. We evaluate performance on bilingual datasets spanning diverse real-world domains by combining our test sets with existing benchmarks, collecting human ratings of rationality, novelty, and engagement, and calibrating an LLM-as-judge against these ratings to enable detailed evaluation and ablations. 
Contributions are as follows:


\begin{figure*}[!ht]
    \centering
    \includegraphics[width=1\linewidth]{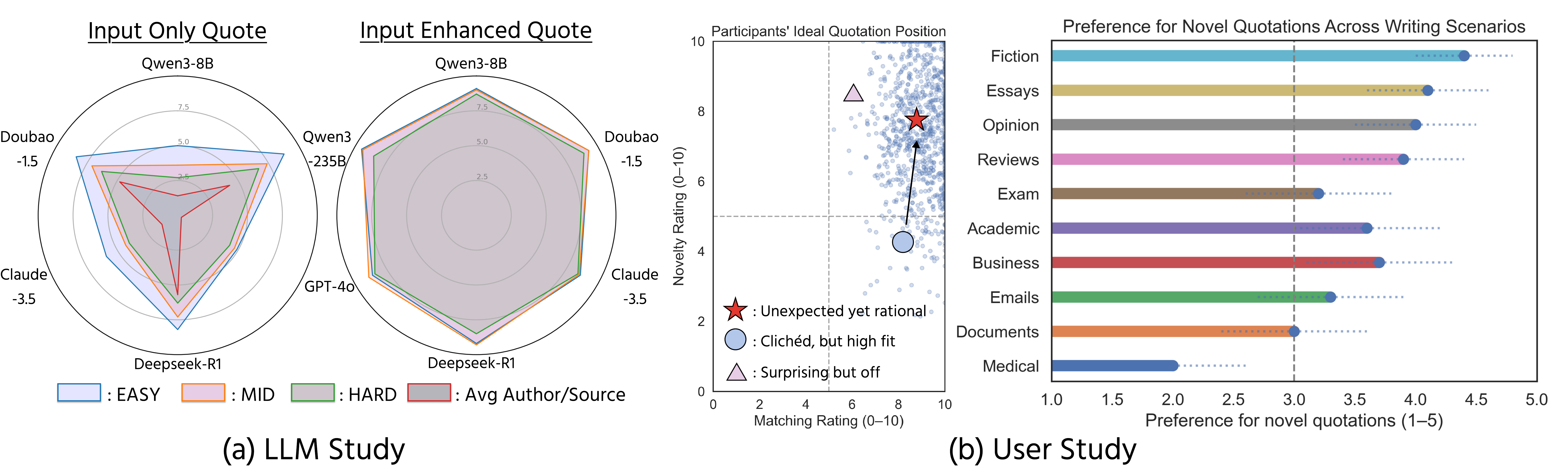}
    \caption{\textbf{Empirical result.} (a) The evaluation results of the only-quote (left) and enhanced-quote (right) scene. All models perform significantly better with enhanced inputs, demonstrating the effectiveness of guided prompt in deep meaning understanding. (b) In user studies, (left) participants perceive ideal quotations as \emph{``unexpected yet rational''}(\includegraphics[height=2ex]{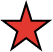}), while current models tend to produce clich\'ed-but-high-fit ones (\includegraphics[height=1.5ex]{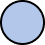}); (right) across various writing scenarios, \emph{novelty} consistently emerges as a key dimension of quotation quality.}
    \label{fig:exp1}
\end{figure*}





\noindent$\bullet$ \, We formalize ideal recommendation as selecting quotes that are unexpected yet rational, grounded in \emph{defamiliarization} and user studies.

\vspace{0.1em}

 \noindent$\bullet$ \, We develop \textsc{NovelQR}, an end-to-end novelty-driven system with a generative label agent that constructs a deep-meaning knowledge base and enables semantic similarity retrieval with fine-grained hard filtering for rationality.

\vspace{0.1em}

\noindent$\bullet$ \, We identify an \emph{auto-regressive continuation bias} in logit-based novelty estimation and propose a token-level method that focuses on ``novelty tokens'' to substantially mitigate this bias.

\section{Related Work}
\noindent \textbf{Quotation Recommendation. }
Work on quotation (quote) recommendation has mainly targeted semantic relevance.
Early methods framed it as learning to rank with handcrafted features~\cite{Quote,10.1145/2911451.2914734,new1,new2,new3,new4,new5}, later replaced by neural models based on CNN/LSTM, Transformers, GRUs, and BERT.
More recently, QUILL~\cite{xiao2025quillquotationgenerationenhancement} adopts a RAG-style framework and offers a comprehensive benchmark.
QuoteR~\cite{quoter} and QUILL provide bilingual test sets, which we use together with our \textsc{NovelQR-Bench} benchmark.
However, existing systems \textbf{largely optimize relevance} and \textbf{do not explicitly model the aesthetic value} or novelty of quotations.

\vspace{0.3em}
\noindent \textbf{Novelty Estimation. }
Textual novelty has been studied mainly from two angles:
\citet{n2} introduce NoveltyBench and view novelty as answer diversity, while \citet{n3,n1} describe that transformers prefer high-frequency words and reduce output diversity.
For operationalizing novelty or surprise, prior work uses bayesian surprise~\cite{mi,surp}, KL divergence~\cite{kl}, and metrics such as embedding distance~\cite{emb}.
These logit-based approaches perform poorly on quote novelty from \textbf{\emph{auto-regressive continuation bias}}.

\section{Empirical Study}

\subsection{Do LLMs truly understand  quotations?}
\label{sec:llm-understanding}
Most quotation systems either generate quotations with LLMs or retrieve them via embedding-based search, typically operating on the quotation in isolation.
This leaves a central question underexplored: \textbf{to what extent do models actually grasp the deep meanings of quotations, and how can this understanding be improved?}

\vspace{0.3em}
\noindent\textbf{Setup. }
We construct a diagnostic evaluation over quotations from three genres (classical Chinese, modern Chinese, and modern English), each paired with expert-written interpretations of their underlying semantics. Quotations are bucketed into three difficulty bands (EASY, MID, HARD). We evaluate several closed- and open-source LLMs on two tasks: (1) explaining the deep meaning of a quotation and (2) identifying its author or source, under two prompting conditions: \emph{quote only} versus an \emph{enhanced quote} that includes auxiliary contextual information. Details are given in Appendix~\ref{appendix:llm-study}, and evaluation results appear in Figure~\ref{fig:exp1}(a).
\vspace{0.3em}

\noindent\textbf{Findings. }
With only the quote as input, all models perform poorly at capturing deep meanings, regardless of size: even on the EASY subset, GPT-4o’s average score remains below the threshold for high-quality semantic understanding, and author/source identification is similarly weak, indicating \textbf{difficulty understanding deep meanings of quotations}. By contrast, enhanced-quote prompts yield substantial gains, where average scores approach 9.0 even on HARD items, and a smaller Qwen3-8B model matches GPT-4o. These results suggest that LLMs can effectively \textbf{grasp deep meanings when supplemented with auxiliary information}. This motivates enriching the quotation knowledge base with labels before retrieval.
\begin{figure*}[!htbp]
    \centering
    \includegraphics[width=1.0\linewidth]{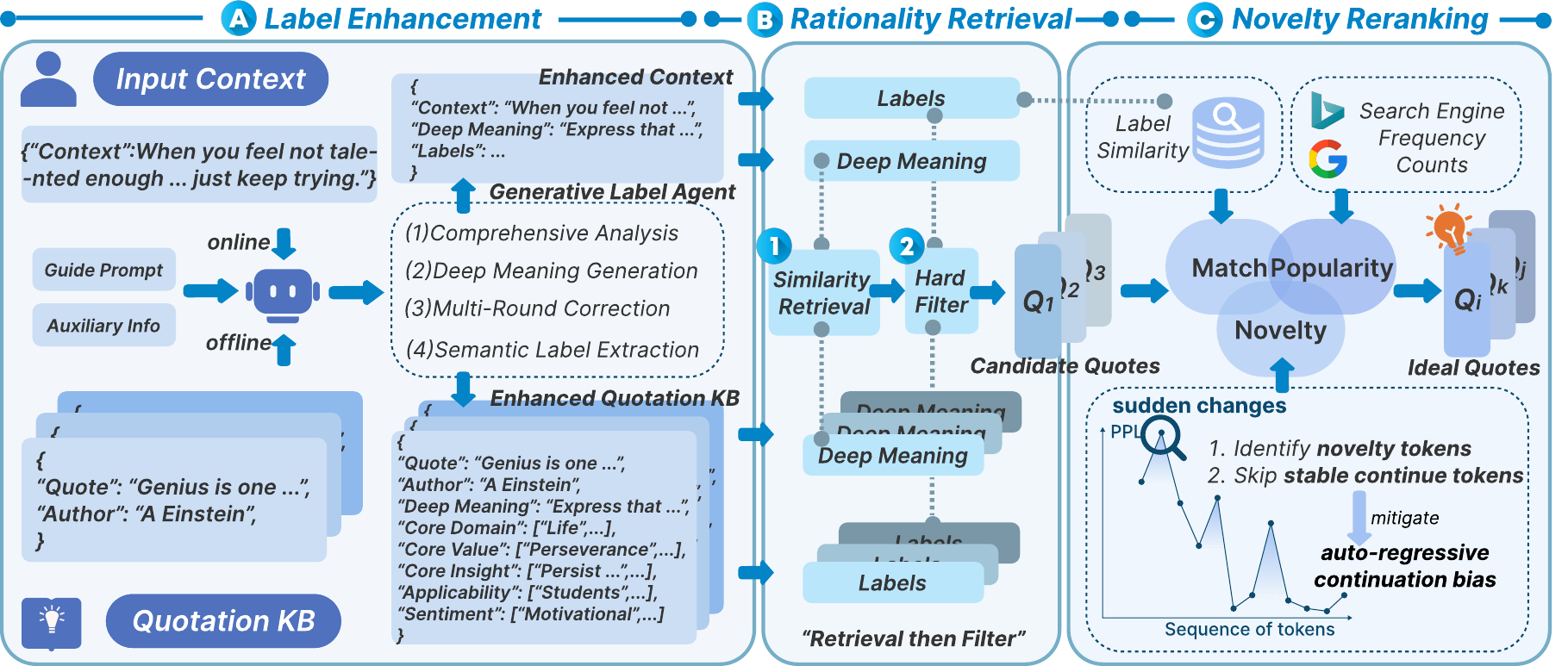}
    \caption{Overview of our novelty-driven quotation recommendation framework: (1) \textbf{Label Enhancement}, where the generative label agent enhances understanding of the quotation knowledge base (KB) and user-given context; (2) \textbf{Rationality Retrieval}, which ``retrieves then filters'' quotations using deep meanings and labels; and (3) \textbf{Novelty Reranking}, which highlights the continuation bias, and introduces the method to mitigate it and estimate novelty.}
    \label{fig:method}
\end{figure*}


\subsection{Do users actually want ``unexpected yet rational'' quotations?}
To ensure that our objective is aligned with user needs, we conduct four complementary user studies (details in Appendix~\ref{appendix:user-study}).

\vspace{0.3em}
\noindent\textbf{Questionnaire. }
We first ran an online questionnaire with $N=964$ respondents across diverse ages and work fields.
On 0--10 scales, an ``ideal'' quotation is rated as almost obligatorily appropriate ($9.1$) and also novel ($7.4$), and users see these two dimensions as complementary rather than conflicting: most place their ideal quotation in the high-match, non-trivially-novel region and many are willing to trade a small amount of fit for extra novelty.
Scenario questions further show that novelty is strongly valued in everyday expressive writing, and open-ended answers repeatedly describe good quotations as those that \emph{``fit the context but still feel fresh''}.

\vspace{0.2em}
\noindent\textbf{Controlled experiments. }
We then ran small controlled studies with 100 participants to test these preferences in behavior.
In rating, pairwise-choice, and cloze-style fill-in tasks that explicitly control contextual fit, participants consistently prefer quotations that are \emph{novel-but-rational} over clich\'ed ones. After reading a short description of a defamiliarization-like effect, they describe it as exactly the kind of impact they want quotations to have in expressive writing, supporting our choice of \emph{unexpected yet rational} as the target objective. As summarized in Figure~\ref{fig:exp1} (b), users perceive an ideal quotation as
\textbf{unexpected yet rational}, which emerges as one important dimension of quotation quality alongside basic appropriateness.

\section{Methodology}

\label{sec:method}
To address the two challenges of ``difficulty understanding deep meanings of quotations'' and ``semantically rational but lacking novelty'', we propose a quotation recommendation system (Figure~\ref{fig:method}), which consists of three steps:

\subsection{Step 1: Label Enhancement}
\label{sec:label-enhancement}
Existing quotation recommendation systems typically retrieve candidates by embedding the \emph{raw quotation text}. However, our empirical analysis shows that even strong LLMs often fail to capture the \emph{deep semantic meanings} of quotations from their surface strings alone, making direct retrieval over raw quotes unreliable. We therefore preprocess our quotation knowledge base (KB) with \textbf{Label Enhancement} before retrieval: \textbf{a generative label agent} produces both deep semantic interpretations and multi-dimensional labels, and performs the same procedure online for user-provided contexts. Following Section~\ref{sec:llm-understanding}, we adopt \emph{Qwen3-8B}~\citep{qwen3technicalreport} as the backbone of this agent.

As illustrated in Figure~\ref{fig:method} (details in Appendix~\ref{appendix: label-agent}), the label agent executes four steps:


   \noindent (1) \textbf{Comprehensive Analysis:} Given auxiliary information (author, source, and related context), the LLM analyzes the quotation from multiple perspectives, including author background, historical–cultural context, and emotional connotations.
    
    \noindent (2) \textbf{Deep Meaning Generation:} Based on the comprehensive analysis beyond the quotation, we extract and concisely summarize the deep semantic meanings within 50 words (``\emph{Express that ...}'').
    
    \noindent (3) \textbf{Multi-round Correction:} 
    The agent self-critiques and refines its explanations for up to $R=3$ rounds, checking for superficiality, over-interpretation, and logical gaps; around 4.6\% of outputs are rejected (protocol in Appendix~\ref{appendix:label-agent-multi-round}).
    
    
    \noindent (4) \textbf{Semantic Label Extraction:} 
    Finally, structured semantic labels are extracted from Core Domains, Insights, Values, Applicability, and Sentiment Tone (Prompts in Appendix~\ref{appendix:prompt-label-agent}).
    
    Through label enhancement, quotations in the KB are mapped into an interpretable deep-meaning space equipped with rich labels, forming the \textbf{basis} for our label-enhanced retrieval module. Retrieving over these interpretations, rather than raw quotation embeddings, yields \textbf{more rational and controllable recommendations}.


\subsection{Step 2: Rationality Retrieval}
\label{Rationality Retrieval}


Traditional systems usually retrieve quotations by measuring similarity over the \emph{raw quotation text}, which we refer to as \textbf{Quote-based Retrieval (QR)}. While embedding-based similarity over surface strings can work for simple, explicit quotations, it often returns candidates that are only superficially related but misaligned with the deeper intent of the context. Moreover, QR skips the interpretive step and does not mimic how humans first analyze a context before choosing an appropriate quotation.

Building on label enhancement, we instead perform retrieval over \emph{deep semantic meanings}, which act as a bridge for semantic retrieval. We term this module \textbf{Label-enhanced Retrieval (LR)}. LR follows a ``\textbf{retrieve–then–filter}'' pipeline.

In the retrieval step, an embedding model encodes the deep meanings of all quotations, as well as  the input context, and we retrieve $TopN$ candidates with the highest similarity in this deep-meaning space.
We then apply a hard filter based on label similarity in the ``Core Domain/Value/Insight'' dimensions with a threshold $T$. Human verification shows that our generated labels have less than 3\% distortion (Appendix~\ref{appendix:human-evaluation-of-deep-meaning-and-labels}), so this signal allows the filter to reliably remove semantically implausible candidates while rarely discarding valid ones. 

We tune on a held-out validation set and use ${TopN=50, T=0.7}$ in all experiments (Appendix~\ref{appendix:lable-retrieval}).
However, the goal of this stage is not to produce the final recommendation, but to \textbf{construct a pool of semantically rational candidates} for the subsequent novelty-aware reranker.

\subsection{Step 3: Novelty Reranking}
\label{sec:novelty-reranking}
Given a candidate pool that is largely rational with respect to the context, the final stage focuses on ranking quotations by their degree of \emph{``unexpectedness''} while mitigating \emph{auto-regressive continuation bias} in standard surprisal-style scores. We combine three factors: \textbf{novelty} \(S_N\), \textbf{semantic match} \(S_M\), and \textbf{popularity} \(S_P\).



\vspace{0.5em}
\noindent\textbf{Novelty.}
We first define quotation novelty. Intuitively, a novel quotation is unfamiliar and difficult for the model to predict under the given context~\citep{surp}. We therefore measure novelty via differences in the model’s own logits.
For a candidate quotation $q = \{x_1, \ldots, x_T\}$, let
\begin{equation}
    p_{\mathrm{prior}}(x_t) =p(x_t | x_1,\cdots,x_{t-1})=p(x_t | X_{<t}),
\end{equation}
\begin{equation}
    p_{\mathrm{cond}}(x_t) =p(x_t | C,x_1,\cdots,x_{t-1})=p(x_t |C,X_{<t}),
\end{equation}
be the token distributions without and with the external context $c$, respectively. We define the log-probability difference $R_t$ and compute \textbf{online},
\begin{equation}
    R_t = \log p_{\mathrm{prior}}(x_t) - \log p_{\mathrm{cond}}(x_t).
\end{equation}
If $R_t > 0$, the token becomes harder to predict under the context, reflecting the kind of \emph{``sudden turn''} or \emph{``surprise''} that we seek.

However, standard logit-based novelty estimators can exhibit errors that stem from what we term \textbf{auto-regressive continuation bias}\footnote{See Appendix~\ref{appendix:novelty-token} for a detailed discussion of continuation bias, our novelty-token design, and the bias analysis from other novelty estimators.}.
In other words, since the model performs inference through next word prediction, some common expressions exhibit continuity problems. For example, after given context ``\emph{When you feel not talented enough to finish this project. Don't worry about that, just keep trying}'', it is difficult to predict ``\emph{Genius is one percent}'' in the beginning, whereas predicting the subsequent phrase ``\emph{inspiration and ninety-nine percent perspiration}” becomes inevitable. If we
\textbf{predict at the word-level or quote-level}, it will \textbf{cause bias} in the final average calculation. To mitigate this, we model quotation novelty at the token level and emphasize \textbf{novelty tokens} rather than treating all tokens uniformly (the bias is illustrated in Figure~\ref{fig:ppl}).

Concretely, we first run the quotation through the language model without context and compute a token-level self-perplexity sequence $\mathrm{PPL}_t = \exp\big(-\log p(x_t \mid x_{<t})\big)$ \textbf{offline}. We then examine how this sequence evolves by taking first- and second-order differences:
\begin{equation}
    \delta_1(t) = \mathrm{PPL}_t - \mathrm{PPL}_{t-1},\textit{ }
|\delta_2(t)| = |\delta_1(t) - \delta_1(t-1)|.
\end{equation}
Large $|\delta_2(t)|$ indicates a sudden change in the local Self-PPL pattern~\citep{xie2024secondorder,shin2024recurve}. To obtain a smooth, non-negative signal, we define
\begin{equation}
\Delta_2(t) = \log\big(1 + |\delta^{\text{pad}}_2(t)|\big),
\end{equation}
where $\delta^{\text{pad}}_2(t)$ denotes $\delta_2(t)$ with appropriate padding at boundaries. We then normalize $\Delta_2(t)$ within each quotation to obtain weights in $[0,1]$:
\begin{equation}
w_t = \frac{\Delta_2(t) - \min_t \Delta_2(t)}{\max_t \Delta_2(t) - \min_t \Delta_2(t) + \epsilon} \in [0,1],
\end{equation}
where $\epsilon$ is a small constant to avoid division by zero and convert $\{w_t\}$ into a distribution over tokens,
\begin{equation}
\tilde{w}_t = \frac{w_t}{\sum_{j=1}^T w_j},
\end{equation}
so that $\sum_t \tilde{w}_t = 1$. Tokens with large $\tilde{w}_t$ are treated as novelty tokens, while smooth continuation regions receive little weight.

Finally, we define the token-level novelty score as a weighted average of log-probability differences:
\begin{equation}
    \small
S_N = \sum_{t=1}^T \tilde{w}_t \Big[\log p(x_t \mid x_{<t}) - \log p(x_t \mid C, x_{<t})\Big].
\end{equation}
Positive contributions to $S_N$ come mainly from novelty tokens whose predictability drops under the context, while continuation-like segments contribute little, thereby reducing bias.

\noindent\textbf{Popularity. }
\label{sec:popularity}
To avoid spuriously treating extremely rare quotations as ``novel'', we add a web-based popularity signal. For each quote $q$ we query \textbf{Bing}\footnote{https://www.bing.com/} with the exact-phrase query under a depersonalized, region-neutral profile and record the count $c$. Counts are collected at fixed UTC snapshots (2025.02-04) and reported in KB for reproducibility. We then map $c$ to a bounded score
\begin{equation}
    \small
    S_{P}= \dfrac{1}{1+e^{-z}} \text{, where }z=\dfrac{log(1+c)-\mu}{\sigma},
\end{equation}
where $\mu$ = 10.53 and $\sigma$ = 2.21 are estimated from $log(1 + c)$ over all quotations.  It is used as a regularizer to \textbf{downweight overly obscure candidates}. Appendix~\ref{popularity} shows that removing popularity yields a consistent drop, and further analyzes \textbf{alignment with human-perceived familiarity} (Spearman $\rho \approx 0.73$, Fleiss' $\kappa = 0.68$) as well as sensitivity to alternative engines (e.g. \textbf{Google}).

\vspace{0.3em}
\noindent\textbf{Semantic Match. }
Although LR already enforces contextual rationality, we still include a semantic matching term to favor quotations that are more coherent and emotionally consistent with the context. We compute cosine similarity between deep-meaning embeddings of the quotation and context, and rescale it to $[0,1]$:
\begin{equation}
    S_M = \frac{1}{2}\left(\frac{\mathbf{h}_q \cdot \mathbf{h}_c}{\|\mathbf{h}_q\| \,\|\mathbf{h}_c\|} + 1\right),
    \end{equation}
$\mathbf{h}_q, \mathbf{h}_c$ denote the semantic embeddings of the quotation and context, respectively.
\vspace{0.5em}
\renewcommand{\arraystretch}{1.05}
    \definecolor{lightgray}{gray}{0.9} 
    \newcommand{\mytablesize}{\fontsize{7.8}{8.8}\selectfont}

\begin{table*}[!htb]
    \centering
    \mytablesize  
    \begin{tabular}{l ccc  ccc  cccc@{\hspace{5pt}}c@{\hspace{5pt}}c}
    \toprule
    & \multicolumn{3}{ c}{\textbf{QuoteR}} 
    & \multicolumn{3}{ c}{\textbf{QUILL}}
    & \multicolumn{6}{ c}{\textbf{\textsc{NovelQR-Bench}} } \\
    
    \cmidrule(lr){2-4} \cmidrule(lr){5-7} \cmidrule(lr){8-13}
    \textbf{Method} & \textbf{Novelty} & \textbf{Match}  &\textbf{Avg}& \textbf{Novelty} & \textbf{Match} & \textbf{Avg}&  \textbf{Novelty} & \textbf{Match}  &\textbf{Avg}& \textbf{HR}$^{\dagger}$ & \textbf{nDCG}$^{\dagger}$ & \textbf{MRR}$^{\dagger}$ \\
    
    \midrule
    \rowcolor{lightgray} \multicolumn{13}{c}{\textit{Model-based Quotation Generation}} \\
     LLM (GPT-based) &  2.85   & 3.00    &   2.93  &  2.76   &  3.10   & 2.93    &  2.85   & 2.99 & 2.92& $\sim$ & $\sim$ & $\sim$    \\
    QuoteR (Bert-based) & 3.55 & 3.77 & 3.66 & 3.55 & 4.08 & 3.82 & 3.21 & 3.88 & 3.54 &$\sim$ & $\sim$ & $\sim$ \\
        \rowcolor{lightgray} \multicolumn{13}{c}{\textit{Retrieval-augmented Quotation Recommendation}} \\
    
QR + w/o Reranker & 3.59 & 3.93 & 3.76 & 3.46 & 4.04 & 3.80 & 3.14 & 3.99 & 3.57 & 0.35 & 0.26 & 0.24 \\
QUILL             & 3.42 & 3.90 & 3.66 & 3.32 & 4.11 & 3.72 & 3.08 & 4.15 & 3.62& 0.15 & 0.12 & 0.11 \\
LR + w/o Reranker & 3.78 & \underline{3.96} & 3.87 & 3.63 & 4.26 & 3.95 & 3.40 & \underline{4.55} & 3.98 & 0.55 & 0.44 & 0.40 \\
LR + bm25         &   3.64   &  3.95    &   3.80   &   3.60   &   4.30   &   3.98   &  3.40    &  4.52   &  3.96 
& 0.40 & 0.30 & 0.23 \\
LR + Bge-large    &   3.75   &  \textbf{4.00}    &   3.88   &   3.60   &  4.33   & 3.97   & 3.61  &    4.54  & 4.08   
&0.56 & 0.39 & 0.33 \\
LR + Qwen3-Re &  3.85  & 3.90     & \underline{3.88}     & 3.75     &  \underline{4.35}    &4.00      &    3.62  & \textbf{4.58}  & 4.10 
& 0.62 & \underline{0.48} & \underline{0.45} \\
LR + GPT     &\textbf{3.90}     &  3.80    & 3.85     & \underline{3.77}     &  4.25    & 4.01     &   3.75   &   4.50  &\underline{4.12}
 & \underline{0.66} & 0.47 & 0.43 \\
LR + Ours         & \underline{3.88} & 3.86 & \textbf{3.88} & \textbf{3.79} & \textbf{4.38} & \textbf{4.09} & \textbf{3.81} & 4.50 & \textbf{4.16} & \textbf{0.70} &\textbf{0.51} & \textbf{0.45} \\

    \bottomrule
    \end{tabular}
    \caption{Comparison of different methods: (1) \emph{Quote-based Retrieval} (QR) retrieves quotations using only quotation text embeddings,  (2) \emph{Label-enhanced Retrieval} (LR) uses deep-meaning embeddings and label-based filtering. Our method consistently outperforms existing approaches across all three datasets, where bm25~\citep{robertson2009bm25}, Bge-large~\citep{bge_embedding}, Qwen3-Reranker~\citep{zhang2025qwen3embedding} and GPT~\citep{sun2024chatgptgoodsearchinvestigating} are the re-ranking methods.  ($^{\dagger}$: Metrics are statistically significant at the 95\% confidence level)}
    \label{tab:modern_comparison}
\end{table*}

\noindent Finally, the reranking score is as follows:
\begin{equation}
S_{final} =  \lambda_1 \cdot S_{N}+\lambda_2 \cdot S_{P}+\lambda_3 \cdot S_{M}.
\end{equation}
By adjusting $\lambda_i$, we balance novelty against rationality factors to ensure that the recommended quotations are both surprising and acceptable.

\vspace{0.3em}
\noindent\textbf{Computational cost. }
The heavy components are run offline, so that online inference only requires embedding similarity searches and logit-difference, with an average end-to-end latency of about $772.2_{-30.5}^{+431.3}$ ms per query (Appendix~\ref{appendix:efficiency}).



\section{Experiments}
In this section, we aim to answer three questions: (1) whether \textsc{NovelQR} \textbf{improves quote recommendation} over strong baselines, (2) whether label-enhanced retrieval yields a \textbf{more semantically coherent} candidate set than text-based retrieval, and (3) whether token-level novelty estimation \textbf{better captures contextual novelty} by mitigating continuation bias. We further examine the \textbf{consistency} between our evaluation and human judgments.

\subsection{Setup}
\noindent\textbf{Datasets.} We evaluate on three high-quality bilingual test sets of 100 instances each: QuoteR, QUILL, and our proposed test set \textsc{NovelQR-Bench}. Together, these sets cover literary, conversational, and expository writing across diverse real-world domains (e.g. literature, science, philosophy, law, etc.). Construction details and statistics are given in Appendix~\ref{appendix:datasets}.

\vspace{0.3em}
\noindent\textbf{Knowledge Base.} The quotations in the knowledge base are from QUILL~\citep{xiao2025quillquotationgenerationenhancement}, which we richly label and embed using the \emph{ACGE text embedding} model~\citep{acge}.

\vspace{0.3em}
\noindent\textbf{Metrics.} 
Our primary retrieval metrics (HR@5, nDCG@5, MRR@5; $^{\dagger}$Statistical significance paired bootstrap testing details in  Appendix~\ref{appendix:significance}) are computed from human-annotated labels (Appendix~\ref{appendix: annotation-novelqr}). In addition, to obtain 1–5 auxiliary scores for Match and Novelty at scale, which are averaged over three random seeds for stability, we use an LLM-as-judge (GPT-4o~\citep{openai2024gpt4ocard}) calibrating against expert ratings (Section~\ref{sec:human_alignment}).


\vspace{0.3em}
\noindent\textbf{Settings.} All methods share the same hyperparameters. Label-enhanced retrieval uses $TopN=50$ and $T=0.7$ (Appendix~\ref{appendix:lable-retrieval}), and the reranking weights are fixed to $\{\lambda_1=0.70, \lambda_2=0.20, \lambda_3=0.10\}$ which tuned on a held-out set over different weight combinations (see Appendix~\ref{appendix:reranking_parameters}, Table~\ref{tab:reranking_parameters}).

\subsection{Main Result}
As shown in Table~\ref{tab:modern_comparison}, \emph{Model-based Quotation Generation} baselines perform worst, mainly due to hallucinations~\citep{xiao2025quillquotationgenerationenhancement} and low appropriateness. \emph{Retrieval-augmented Quotation Recommendation} methods perform substantially better, especially those based on semantic matching. Within this family, moving from \emph{Quote-based Retrieval} (QR+w/oReranker) to our \emph{Label-enhanced Retrieval} (LR+w/oReranker) yields a large gain in Match (from 3.99 to 4.55), indicating that the first-stage rationality retrieval provides a much \textbf{stronger candidate set} for subsequent reranking.

Fixing LR as the retriever, we then compare different rerankers. Across all test sets, our novelty-aware reranker (LR+Ours) \textbf{achieves the best overall performance}, substantially boosting novelty while maintaining high match and strong ranking metrics. Additionally, in a human multiple-choice study 
 (Appendix~\ref{appendix:user-study-study2}), \textbf{78\%} of selections favor our system. Improvements hold for classical literary quotations, modern conversational contexts, and expository writing, suggesting that our framework is \textbf{not restricted to a single genre}.
\renewcommand{\arraystretch}{1.10}
\definecolor{lightgray}{gray}{0.9} 

\renewcommand{\arraystretch}{1.0}
\begin{table}[!htb]
    \centering
    \mytablesize  
    \begin{tabular}{l @{\hspace{6.0pt}}c@{\hspace{5.5pt}}c@{\hspace{5.5pt}}c@{\hspace{5.5pt}}c@{\hspace{5.5pt}}c}
    \toprule
    & \multicolumn{5}{c}{\textbf{\textsc{NovelQR-Bench}}} 
\\
    
    \cmidrule(lr){2-6}
    \textbf{Method}&\textbf{Novelty}& \textbf{Match} & \textbf{HR}$^{\dagger}$ & \textbf{nDCG}$^{\dagger}$ & \textbf{MRR}$^{\dagger}$  \\
    
    \midrule
Self-BLEU   &3.55 &  4.48&  0.50&0.39 & 0.37     \\
Embedding-Dis   & 3.66&\textbf{4.56}    & 0.50 & 0.41&0.37      \\
Surprisal  &3.66 & 4.31&    0.55 &0.44 &0.40      \\
\, + \emph{Novelty-token} & 3.73 & 4.39 & 0.62 & 0.45 & 0.42 \\
KL-Div  &3.48 &4.39 & 0.61 &0.43 & 0.37      \\
\, + \emph{Novelty-token} & 3.64 & 4.40 & 0.61 & 0.45 & 0.40 \\
 Uniform Avg & 3.66 & 4.45 & 0.63 & 0.46 & 0.41 \\
 TopK Avg    & 3.68 & 4.48 & 0.65 & 0.47 & 0.42 \\

\rowcolor{lightgray} \multicolumn{6}{l}{\emph{Ours (Novelty-token)}}     \\

\quad Qwen3-8B       & \textbf{3.81}& \underline{4.50}   & \underline{0.70} & \textbf{0.51} & \textbf{0.45}      \\
\quad Qwen3-0.6B       & 3.74&   4.46        & 0.66 & 0.48 & 0.42      \\
\quad Qwen3-32B    &\underline{3.77}    & 4.45        & \textbf{0.71} & \underline{0.50} & 0.44      \\
\quad Qwen2.5-7B      & 3.72& 4.42           & 0.65 & 0.47 & 0.41      \\
\quad Llama3-8B       & 3.66& 4.38            & 0.61 & 0.43    & 0.38         \\
\quad GLM3-6B         & 3.75& 4.44      & 0.66  &0.45    & \underline{0.44}        \\


    \bottomrule
    \end{tabular}
    \caption{Evaluation results of various methods for novelty estimation. The other methods are implemented by \emph{Qwen3-8B} model and \emph{ACGE} text embedding model. \\
    ($^{\dagger}$: Statistically significant with 95\% confidence)}
    \label{tab:ab3}
\end{table}


    
    


\subsection{Ablation Study}
\noindent\textbf{Novelty Estimation. }
\label{sec:novelty-estimation}
Building on our token analysis in Appendix~\ref{appendix:continuation-bias-analysis}, we observe that likelihood-based metrics such as Surprisal and KL-Divergence are systematically distorted by the \emph{auto-regressive continuation bias} while Embedding Distance and Self-BLEU are not. To further validate the effectiveness of our \emph{novelty-token}, we compare our method against several used alternatives, their variants equipped with novelty-token weighting (+ \emph{Novelty-token}), and two token-level ablations of our method: a uniform average over token-wise logit gaps and a $TopK$ variant. We use each as a drop-in replacement for $S_N$ and also test different LLMs (Detailed formulations in Appendix~\ref{appendix:definition-of-other-novelty-estimation-method}).

As shown in Table~\ref{tab:ab3}, existing metrics and the two token-level ablations fail to accurately capture contextual novelty. When we equip logit-based baselines with our novelty-token weighting, their performance improves consistently, demonstrating that our novelty-token design is both \textbf{effective and well suited} to this setting, although our full estimator $S_N$ still \textbf{achieves the best overall performance}. Moreover, analyses across different model sizes and families show minimal performance variation, indicating that our estimator \textbf{remains robust} even with relatively small models.

\vspace{0.5em}
\renewcommand{\arraystretch}{1.10}
\definecolor{lightgray}{gray}{0.9} 

\renewcommand{\arraystretch}{1.0}
\begin{table}
    \centering
    \mytablesize 
    \begin{tabular}{lc@{\hspace{10.0pt}}c}
        \toprule
       \textbf{Retrieval setting} & \textbf{Match} & \textbf{$\Delta$(+)}   \\
        \midrule
        \rowcolor{lightgray} \multicolumn{3}{c}{\emph{Backbone comparison}} \\
        Quote-only embeddings (QR)                & 4.15 & $\sim$ \\
        Deep-meaning embeddings only              & \underline{4.45} & 0.30 \\
        Label-only embeddings  & 4.25 & 0.10 \\
        Deep-meaning  labels (LR)    &  \textbf{4.50} & 0.35 \\
        \midrule
        \rowcolor{lightgray} \multicolumn{3}{c}{\emph{Label-filter variants (with deep-meaning retrieval)}} \\
        No label filter                &  4.39  & $\sim$ \\
        Domain only                    & 4.44  & 0.05 \\
        Value only                     & 4.45  & 0.06 \\
        Insight only                   & 4.45  & 0.06 \\
        Domain + Value                 & 4.47  & 0.08 \\
        Domain + Insight               & 4.45  & 0.06 \\
        Value + Insight                & \underline{4.48}  & 0.09 \\
        Domain + Value + Insight       & \textbf{4.50}  & 0.11 \\
        \bottomrule
      \end{tabular}
      \caption{Effect of deep-meaning retrieval and label-based filtering on Match.}
      \label{tab:label-ablation}
\end{table}

\begin{figure}[ht]
    \centering
    \includegraphics[width=0.90\linewidth]{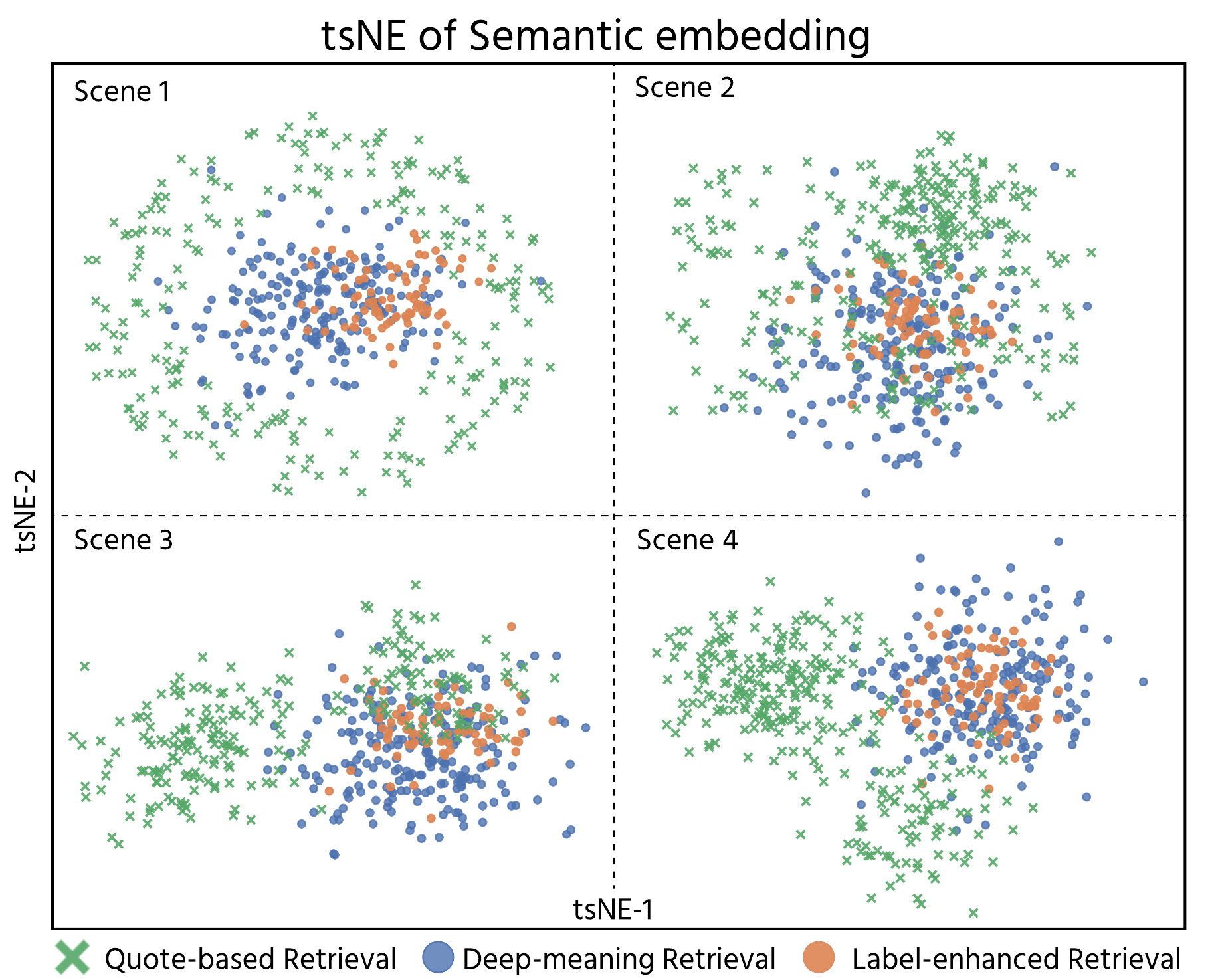}
    \caption{Semantic embedding visualization (T-SNE) of retrieved quotations using different methods. \emph{Label-enhanced} shows tighter clustering and better semantic consistency compared to \emph{Quote-based retrieval}. }
    \label{fig:tsne}
\end{figure}
\begin{figure}
    \centering
    \includegraphics[width=1\linewidth]{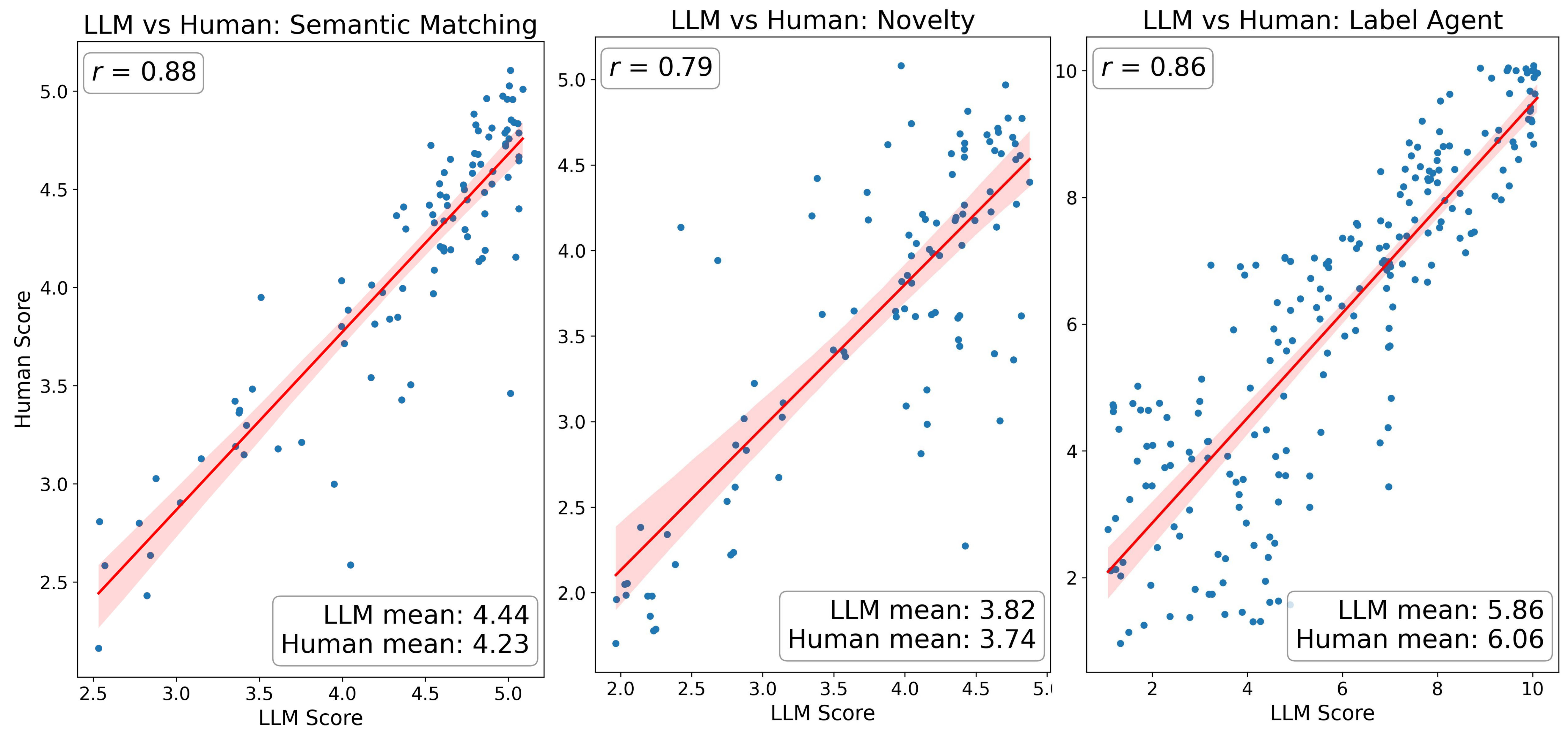}
    \caption{The Correlation between our LLM-as-judge evaluation and human scores. To avoid overlapping points, random jitters were added to ratings. }
    \label{fig:human}
\end{figure}
\noindent\textbf{Effect of LLM-Based Labels. }
We next ask whether deep-meaning retrieval and label filtering are necessary. We compare four retrieval settings: (1) retrieving quotations using only quotation text embeddings (QR), (2) using deep-meaning embeddings without any label filter, (3) retrieving embeddings of the concatenated \emph{Core Domain}, \emph{Value}, and \emph{Insight} labels, and (4) our full setting (LR), which combines deep-meaning retrieval with label-based filtering on these three dimensions. We also fix deep-meaning retrieval and vary which label subsets are used in the filter to examine the contribution of each dimension (Table~\ref{tab:label-ablation}).

Compared with quote-based embeddings, deep-meaning embeddings already yield better semantic performance. Adding label-based filtering further improves results, demonstrating the \textbf{effectiveness} of the label filter. Moreover, performance remains stable across different label variants, suggesting that the label filter is \textbf{robust} to each dimension.


\vspace{0.5em}
\noindent\textbf{Semantic Structure. }
To assess the semantic quality of our retrieval module, we compare \emph{Label-enhanced Retrieval} (LR) with \emph{Quote-based Retrieval} (QR) by visualizing the embeddings using t-SNE (Figure~\ref{fig:tsne}). QR, which retrieves directly from raw quotation text, yields scattered and mixed clusters, indicating weaker semantic coherence. In contrast, LR produces more coherent, contextually aligned clusters, suggesting that our method \textbf{captures the underlying semantics more faithfully}.



\subsection{Human Alignment}
\label{sec:human_alignment}
Given the subjectivity, we randomly sample 500 instances and collect 1–5 ratings from three literature experts (ICC = 0.81 for Match, 0.76 for Novelty; Appendix~\ref{appendix: annotation-match-novelty}). Figure~\ref{fig:human} compares human and LLM-as-judge scores, showing \textbf{general alignment} ($\rho > 0.79$) between model and human while designed prompts and criteria make the LLM-as-judge framework a reasonably reliable proxy for human judgments. In Appendix~\ref{appendix:robustness}, we further confirm the \textbf{stability} across different LLM judges (GPT-4o, Claude 3.5, Gemini-1.5 and Qwen-Plus) and sampling temperatures of 0 and 0.7. 
\begin{figure*}[t]
    \centering
\includegraphics[width=1.00\linewidth]{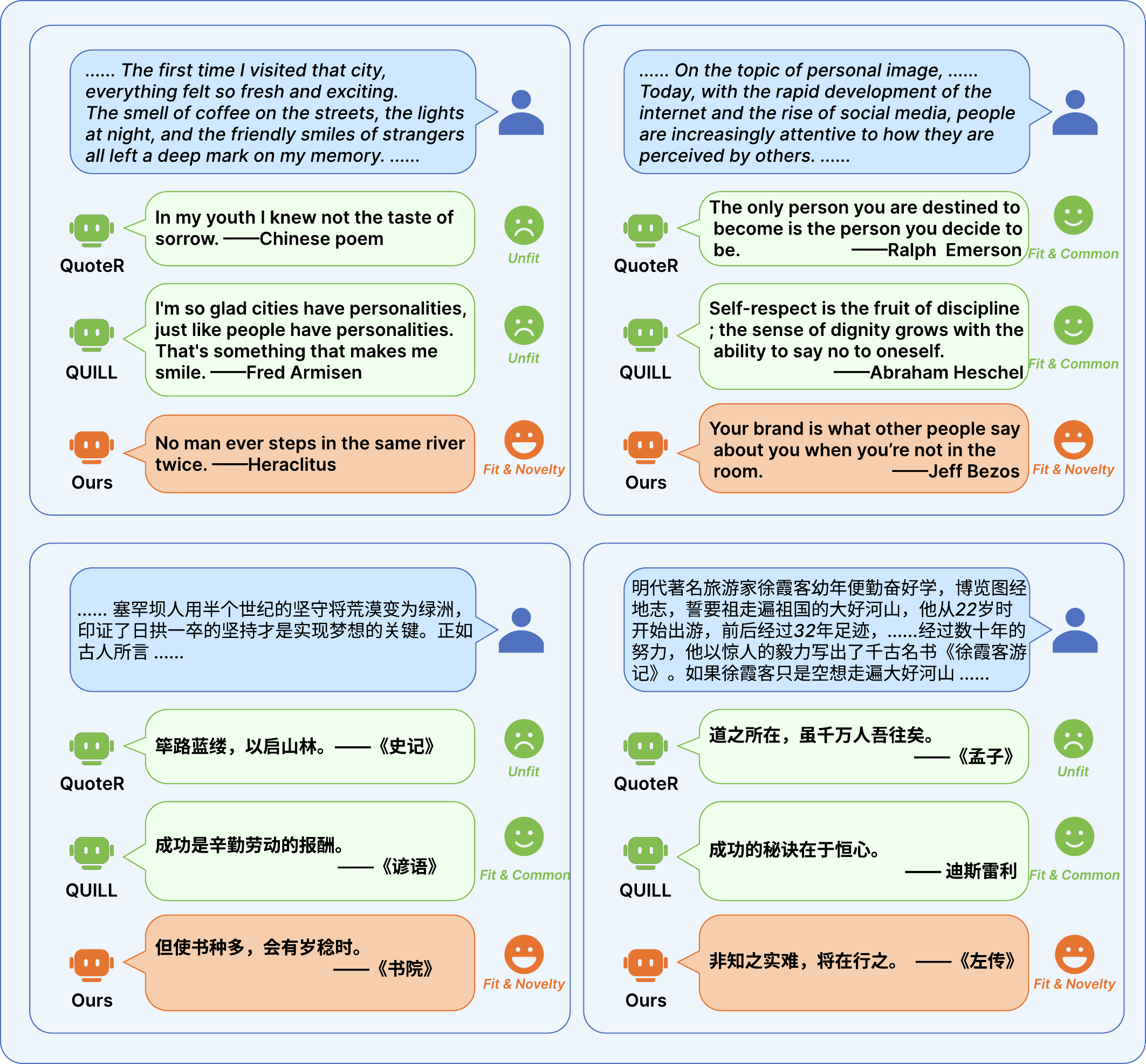}
    \caption{More cases of recommendation. \textbf{Our method can recommend more in-depth citations, rather than just semantically relevant ones.}}
    \label{fig:case-study}
\end{figure*}

\subsection{Practicality, Cost, and Scalability}
Our pipeline separates a one-time offline indexing cost from lightweight online inference. Offline, we construct a labeled quotation knowledge base by generating deep-meaning explanations and structured labels for the quotation pool. This preprocessing is performed once and the resulting labeled KB can be reused across future queries. Online, the system only requires a single label-generation step for the input context, followed by embedding retrieval, label-based filtering, and token-level novelty scoring. Therefore, the online cost is comparable to a standard retrieval-augmented pipeline rather than requiring expensive per-candidate generation. In our implementation, \textbf{the additional cost mainly comes from the offline KB construction, while the online latency remains practical for interactive use.} Moreover, scalability is favorable: enlarging the quotation pool increases only the offline preprocessing cost \textbf{approximately linearly}, while online serving can still rely on standard retrieval and batched reranking over a small candidate set. More details about efficiency and scalability are provided in Appendix~\ref{appendix:efficiency}.


\section{Case Study}
To provide a more intuitive illustration, we present examples from the experiments and compare our results with those from QuoteR and QUILL in Figure~\ref{fig:case-study}. In these cases, the baseline systems are easily \textbf{misled by surface-level cues} (e.g., interpreting a passage about \emph{longing when returning to the city} as simply being \emph{a city}, or the classical theme of \emph{Autumn Thoughts} as merely \emph{autumn}) and therefore fail to recommend an ideal quote, while our method tracks the deeper intent of the context. This highlights that \textbf{capturing deep meanings is essential}. 


\section{Conclusion}
From our large-scale user studies, we presented a defamiliarization-inspired quotation recommendation framework \textsc{NovelQR} that targets quotations which are ``unexpected yet rational''. Methodologically, we propose a logit-based, token-level novelty estimator that mitigates the \emph{auto-regressive continuation bias}. Experiments on multi-genre data with both human and LLM-as-judge evaluation suggest that our system can recommend quotations that are more appropriate and more novel, showing its potential as a practical writing assistant.

\section*{Limitations}
Yet, as Shakespeare noted, \emph{``There are a thousand Hamlets in a thousand people's eyes''}—novelty is inherently subjective and varies among individuals. While our human-anchored, LLM-based estimation provides a practical proxy, it still cannot fully capture such subjectivity; developing more comprehensive and robust evaluation frameworks for novelty remains important future work.

\section*{Acknowledgements}
We thank the anonymous reviewers and area chairs for their thoughtful and constructive feedback, which helped improve this paper. We are also sincerely grateful to our teachers for their guidance and support throughout this work.

\bibliography{custom}

\appendix
\clearpage
\newpage

\section*{Appendix}

\section{Result of Reranking Parameters $\lambda_i$}
\label{appendix:reranking_parameters}
Here we present the table from the ablation study section titled ``Impact of Reranking Score Parameters''.
\renewcommand{\arraystretch}{1.00}
\definecolor{lightgray}{gray}{0.9} 

\begin{table}[!htb]
    \centering
    \small  
    \begin{tabular}{ccc ccc}
    \toprule
\multicolumn{3}{c}{\textbf{Parameters}} 
    & \multicolumn{3}{ c}{\textbf{LLM-as-Judge}}
\\
    
    \cmidrule(lr){1-3} \cmidrule(lr){4-6} 
    \textbf{$S_N$} & \textbf{$S_P$} & \textbf{$S_M$}  & \textbf{Novelty} & \textbf{Match}  & \textbf{Avg}\\
    
    \midrule
1.00 & 0.00 & 0.00 & \textbf{3.82} & 4.41 & 4.115 \\
0.70 & 0.30 & 0.00 & 3.79 & 4.47 & 4.130 \\
0.50 & 0.50 & 0.00 & 3.69 & 4.46 & 4.075 \\
0.70 & 0.00 & 0.30 & 3.71 & 4.46 & 4.085 \\
0.70 & 0.15 & 0.15 & 3.80 & \underline{4.50} & \underline{4.150} \\
\rowcolor{lightgray} 0.70 & 0.20 & 0.10 & \underline{3.81} & \textbf{4.50} &\textbf{ 4.155} \\
0.50 & 0.25 & 0.25 & 3.72 & 4.47 & 4.095 \\


    \bottomrule
    \end{tabular}
    \caption{Performance under different weight combinations of novelty ($S_n$), popularity ($S_p$), and semantic matching ($S_m$). (statistically significant at $p<0.05$). }
    \label{tab:reranking_parameters}
\end{table}
Overall, as shown in Table~\ref{tab:reranking_parameters}, when the novelty score remains the dominant component, the overall score fluctuates but consistently achieves good performance. Therefore, the final combination $\{S_N=0.70, S_P=0.20, S_M=0.10\}$ is selected based on \textbf{held-out tuning}. Since real-world scenarios do not uniformly prefer high novelty (Figure~\ref{fig:exp1}), a writing assistant can adjust the weighting parameter~$\lambda$ to adapt the balance accordingly.

\section{Ablation Studies on Label-based Retrieval Method}
\label{appendix:lable-retrieval}
In our reranking system, its effectiveness relies on the assumption that the candidate quotations themselves are semantically reasonable. 
Therefore, in this experiment, we aim to verify the semantic quality of the label-based retrieval approach as well as the effect of the parameter settings used in this retrieval process. 
Unlike direct quote-based retrieval from the entire corpus, this approach retrieves and filters candidates based on generative labels and deep semantic meanings. 
Specifically, in label-based retrieval we set the number of top retrieved items for deep semantic matching as 
\[
TopN = \{50, 100, 150, 200\}
\]
and the semantic threshold for hard filtering based on labels as
\[
T = \{0.5, 0.7, 0.9\}.
\]
We then use an LLM-as-judge to evaluate the semantic alignment between each quotation and its context as the effectiveness metric.

From the experimental results in Table~\ref{tab:label_retrieval}, we observe that increasing \(TopN\) does not improve the semantic alignment score. 
Therefore, we choose \(TopN=50\) for faster response. 
Although increasing the threshold improves alignment scores, the number of quotations that remain after filtering becomes fewer than five, resulting in too few candidates. 
\textbf{Consequently, we finally select \(T=0.7\), which yields an average semantic alignment score of 4.5, and use these parameters for semantic retrieval.} 
Furthermore, results from the main experiments also show that label-based retrieval achieves \textbf{higher semantic alignment scores} compared with direct quote-based retrieval, which validates the effectiveness of our method and supports our underlying assumption.
\renewcommand{\arraystretch}{1.10}
\definecolor{lightgray}{gray}{0.9} 

\renewcommand{\arraystretch}{1.0}
\begin{table}[!htb]
    \centering
    \small  
    \begin{tabular}{ccccc}
\toprule
\textbf{TopN}       & \textbf{T} & \textbf{Row Quote}           & \textbf{Final Quote} & \textbf{Length} \\
\midrule
 & 0.5 & & 4.3 & 46.7 \\
\rowcolor{lightgray}50 & 0.7 &   4.2                   & 4.5 & 18.0 \\
                     & 0.9 &                      & 4.7 & 1.3  \\
\midrule \multirow{3}{*}{100} & 0.5       & \multirow{3}{*}{4.1} & 4.0          & 91.5   \\
                     & 0.7       &                      & 4.5          & 30.4   \\
                     & 0.9       &                      & 4.6          & 3.2    \\
\midrule\multirow{3}{*}{150} & 0.5       & \multirow{3}{*}{4.1} & 4.2          & 136    \\
                     & 0.7       &                      & 4.2          & 46.1   \\
                     & 0.9       &                      & 4.6          & 3.5    \\
\midrule\multirow{3}{*}{200} & 0.5       & \multirow{3}{*}{4.0} & 4.0          & 180    \\
                     & 0.7       &                      & 4.3          & 32.2   \\
                     & 0.9       &                      & 4.5          & 3.7    \\    
    \bottomrule
    \end{tabular}
    \caption{Ablation Study of Label retrieval. The result shows that selecting the parameters $\{TopN=50,T=0.7\}$ for label-based retrieval \textbf{achieves the best semantic alignment score}.}
    \label{tab:label_retrieval}
\end{table}

\section{Ablation Studies on Popularity}
\label{popularity}
\subsection{Effect on performance}
In Section~\ref{sec:popularity}, we study the impact of the web-based popularity score $S_P$ on our system by comparing: (1) \textbf{w/o popularity}, which drops $S_P$ and relies only on semantic match and token-level novelty; and (2) \textbf{Bing}\footnote{https://www.bing.com/}, \textbf{Google}\footnote{https://www.google.com/}, and \textbf{Baidu}\footnote{https://www.baidu.com/}, which use the same procedure in Section~\ref{sec:method} but with different search engines to estimate document frequency. For each variant we recompute $S_P$, rerun the reranking stage, and report HR@5, nDCG@5, MRR@5 and LLM-as-Judge score (Novelty and Matching) on the bilingual test sets. As shown in Table~\ref{tab:pop-performance}, all popularity-enabled variants outperform the w/o-popularity baseline, and the three engines yield very similar trends. This indicates that \textbf{incorporating a coarse web-frequency signal is beneficial} and that our method is not sensitive to the specific choice of search engine.
\renewcommand{\arraystretch}{1.00}
\definecolor{lightgray}{gray}{0.9} 

\begin{table}[!htb]
    \centering
    \small  
    \begin{tabular}{lccccc}
        \toprule
        \textbf{Variant} & \textbf{Novelty} & \textbf{Match} & \textbf{HR} & \textbf{nDCG} & \textbf{MRR} \\
        \midrule
        \textbf{w/o $S_P$}    & 3.70 & 4.46 & 0.66 & 0.48 & 0.42 \\
        \rowcolor{lightgray}\textbf{Bing}      & 3.82 & 4.50 & 0.70 & 0.51 & 0.45 \\
        \textbf{Google}       & 3.80 & 4.49 & 0.69 & 0.50 & 0.44 \\
        \textbf{Baidu}        & 3.79 & 4.47 & 0.68 & 0.49 & 0.43 \\
        \bottomrule
      \end{tabular}
      \caption{Effect of different popularity variants on ranking performance. The result shows that incorporating a web-frequency signal is \textbf{beneficial} and that our method is \textbf{not sensitive} to the specific choice of search engine.}
      \label{tab:pop-performance}
\end{table}
\begin{figure}
    \centering
    \includegraphics[width=0.5\textwidth]{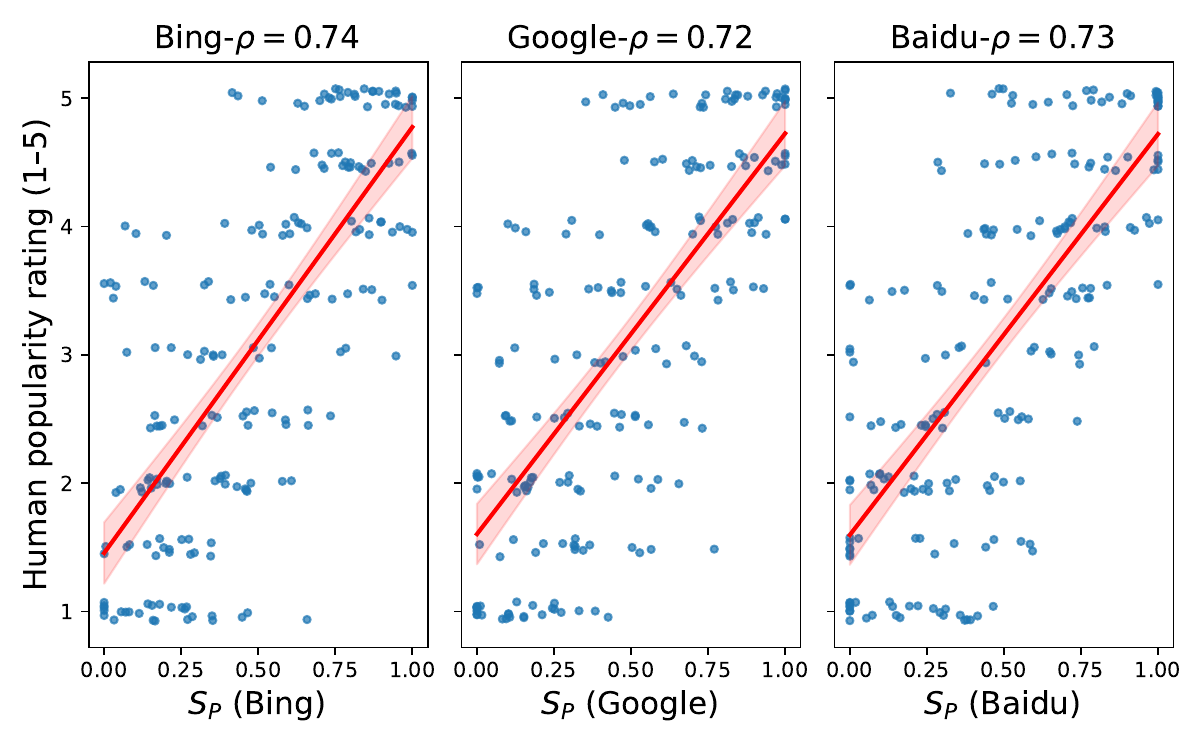}
    \caption{\textbf{Alignment between the web-based popularity score $S_P$ and human-perceived popularity.} The result shows a clear positive relationship between $S_P$ and human judgments, suggesting that our web-based popularity score is a \textbf{reasonable approximation} of perceived quotation popularity. ($\kappa = 0.68$)}
    \label{fig:popularity_human}
\end{figure}

\subsection{Human-perceived popularity alignment}
We also run a small human study to check whether $S_P$ agrees with how people perceive quotation popularity. We sample $N=200$ quotations from the KB, and ask three annotators to rate, on a 1--5 scale, how familiar or widely known each quotation is. We average the human scores and compute the Spearman correlation with $S_P$ obtained from Bing. The resulting correlation Figure~\ref{fig:popularity_human} shows a clear positive relationship between $S_P$ and human judgments, suggesting that our web-based popularity score is a \textbf{reasonable approximation} of perceived quotation popularity and \textbf{suitable as a regularizer} in the final ranking. See Appendix~\ref{appendix: annotation-popularity} for more details ($\kappa = 0.68$).

\section{Datasets}
\label{appendix:datasets}
\subsection{Overview}
Table~\ref{tab:dataset-overview} summarizes the three test sets used in our experiments. Across all three test sets, our system consistently outperforms retrieval and generation baselines. Importantly, the relative gains are stable from canonical literary quotations to modern quotations and to out-of-domain contexts in reports, news, and essays, suggesting that the proposed framework is \textbf{not tailored to a specific genre}.

\renewcommand{\arraystretch}{1.10}
\definecolor{lightgray}{gray}{0.9} 

\renewcommand{\arraystretch}{1.0}
\begin{table}[!htb]
    \centering
    \tiny  
    \begin{tabular}{l@{\hspace{2pt}}ccc}
        \toprule
        \textbf{Dataset} & \textbf{\#Instances}  & \textbf{Main domains} & \textbf{Context style} \\
        \midrule
        \textbf{QuoteR}      & 100  & literature, philosophy & short narrative/expository \\
        \textbf{QUILL}  & 100  & books, interviews, forums & modern, conversational \\
        \textbf{\textsc{NovelQR-Bench}}     & 100  & reports, news, student essays & expository, argumentative \\
        \bottomrule
      \end{tabular}
      \caption{Overview of our three bilingual test sets. Together they cover classical and modern quotations and contexts from \textbf{literary, conversational, and expository writing.}}
      \label{tab:dataset-overview}
\end{table}

\subsection{Construction of \textsc{NovelQR-Bench}}
Existing benchmarks (QuoteR, QUILL) mainly focus on literary and conversational contexts. To better test robustness in more informational and argumentative settings, we construct \textbf{\textsc{NovelQR-Bench} } as follows.

(1) \textbf{Context sampling.} We sample 100 contexts in total from three sources: (i) public reports and opinion pieces\footnote{https://paper.people.com.cn/}, (ii) news articles\footnote{https://www.xinhuanet.com/} (e.g., technology, society, finance), and (iii) high-school and undergraduate essays\footnote{https://www.zuowen.com/} on themes such as persistence, parting, and self-discipline. We filter for contexts with length between 80 and 300 tokens and remove duplicated or near-duplicated passages.

(2) \textbf{Candidate quotations.} For each context, we retrieve the $K=50$ quotations from our bilingual KB using a strong embedding-based retriever (Label-based  and Quote-based retrieval). The retrieved candidates are randomly shuffled before annotation to avoid position bias.

(3) \textbf{Human relevance labels.} Three annotators independently mark up to three quotations per context that they consider ``appropriate and expressive'' for the given passage. We take the union of their selections as the relevant set when computing HR, nDCG, and MRR. No system outputs are shown during annotation, and we only use the raw texts without any personal metadata. (Appendix~\ref{appendix: annotation-novelqr})

\clearpage

\section{User Study}
\label{appendix:user-study}

Here we will provide additional details of our user studies designed to verify that quotation novelty is not merely a philosophical construct, but a user-perceived and optimizable objective in quotation recommendation.
Concretely, we aim to answer three questions that complement the empirical results in the main paper:

(1) \emph{How users conceptualize ``appropriateness'' and ``novelty'' for quotations,}

(2) \emph{Whether they see these as complementary rather than mutually exclusive,}

(3) \emph{How the preference for novel quotations varies across writing scenarios.}

Building on these findings, subsequent studies (reported in later subsections) use controlled choice experiments and utility modeling to connect user attitudes with actual selection behavior.
We first present the full questionnaire used in \textbf{Study~1}, which focuses on users' perceptions and self-reported preferences.

\subsection{Study 1: Perception and Scenario Questionnaire}
\label{appendix:user-study-study1}
This is a questionnaire-based survey.
It consists of five parts:
(A) demographics and writing background,
(B) views on appropriateness and novelty,
(C) direct comparison questions between different types of quotations,
(D) preferences across writing scenarios,
and (E) self-reported behavior and open-ended feedback.
The full instrument is reproduced in Appendix~\ref{appendix:prompt-questionnaire}.

\paragraph{Collection. }
We first analyze responses to the questionnaire. We distributed the survey via Wenjuanxing\footnote{\url{https://www.wjx.cn}}, a widely used online questionnaire platform, and collected \textbf{a total of $N = 964$ completed responses}.
All responses passed our basic attention checks, so we retained all 964 for analysis.

\paragraph{Participants.}
In \textbf{Part A}, we asked participants for their \emph{age group} and \emph{primary work field}.
The sample covers all age groups from 18--24 up to 55+, and spans multiple work fields including education, research, industry, and other professions.
Table~\ref{tab:study1-demographics} summarizes the distribution (\textbf{basically covering users of all categories}).

\renewcommand{\arraystretch}{1.10}
\definecolor{lightgray}{gray}{0.9} 

\renewcommand{\arraystretch}{1.0}
\begin{table}[!htb]
    \centering
    \small  
    \begin{tabular}{l r @{\hspace{14pt}} lr}
        \toprule
        \multicolumn{2}{c}{\textbf{Age group}} & \multicolumn{2}{c}{\textbf{Primary work field}}\\
        \midrule
        18--24 & 218 &Education & 312 \\
        25--34 & 376 &Research & 271 \\
        35--44 & 231 &Industry & 307 \\
        45--54 & 96 &Other & 74 \\
        55+    & 43 & - & - \\
        \bottomrule
    \end{tabular}
    \caption{Summary of participants in Study~1 ($N=964$).}
    \label{tab:study1-demographics}
\end{table}


Most participants reported writing long-form texts at least monthly, and a majority indicated that they use quotations at least occasionally in their writing.
Below we summarize the key quantitative findings relevant to how users perceive and prioritize appropriateness and novelty.

\paragraph{Appropriateness vs.\ Novelty. }
Participants rated the importance of \emph{contextual appropriateness} and \emph{novelty} for an ``ideal'' quotation on 0--10 scales (\textbf{Q6-Q7}).

\begin{figure}[H]
    \centering
\includegraphics[width=1\linewidth]{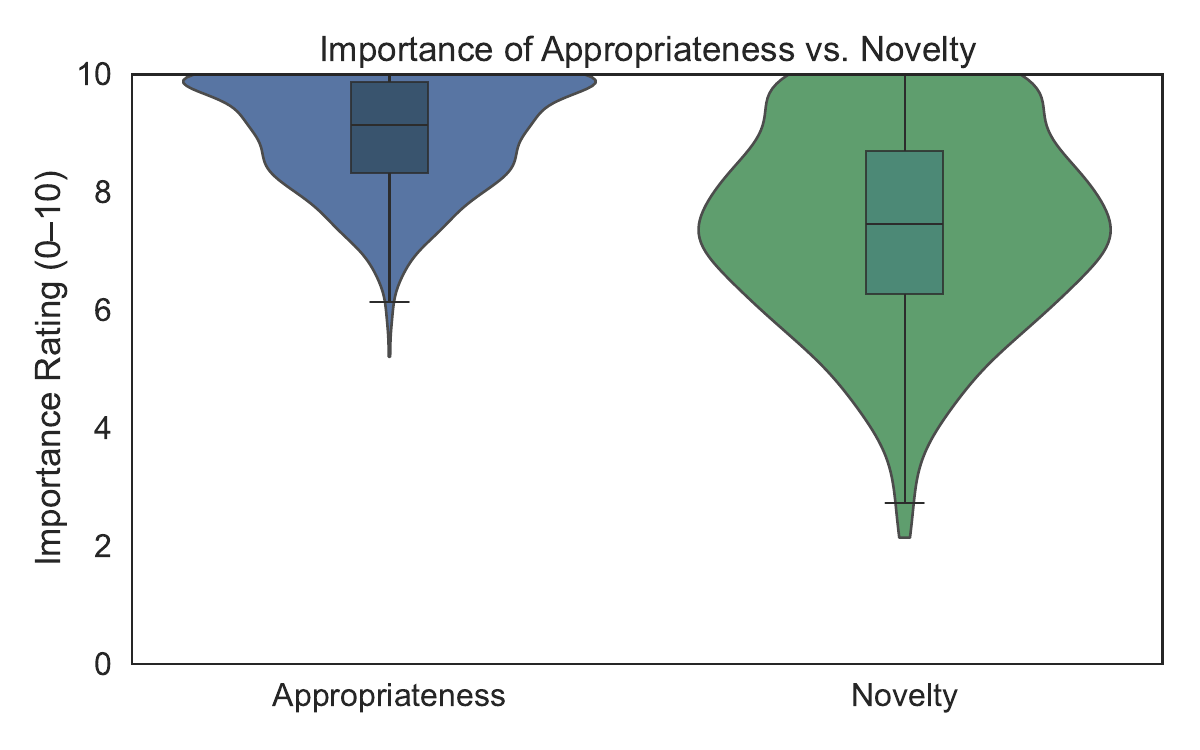}
\caption{Importance ratings (0--10) for appropriateness and novelty in an ``ideal'' quotation (Q6--Q7).
Both are rated highly, with \textbf{appropriateness near-essential and novelty clearly important}.}
\label{fig:study1-importance}
\end{figure}

The mean importance of appropriateness was $9.1$ (SD $1.2$), while the mean importance of novelty was $7.4$ (SD $1.8$), both significantly above the neutral midpoint of 5 (one-sample $t$-tests, $p < 10^{-10}$ for both).
A box plot or violin plot comparing these two distributions (Figure~\ref{fig:study1-importance}) makes the contrast visually clear: respondents almost unanimously \textbf{treat appropriateness as a must-have requirement, and also assign substantial importance to novelty}.

\paragraph{Complementary vs.\ Mutually Exclusive. }
\textbf{Q8} further probes how users conceptually relate the two dimensions through five Likert statements (1 = strongly disagree, 5 = strongly agree), which reports the mean and standard deviation for each statement. 
Respondents strongly agree that appropriateness is a prerequisite (Q8(a)) (Mean = 4.6, SD = 0.7), and they also agree that, given appropriateness, less clich\'ed and more original quotations are preferred (Q8(b)).
They additionally endorse the two-dimensional view in Q8(c).
In contrast, they clearly reject the two extreme views in Q8(d) and Q8(e) (Mean = 1.8, SD = 0.9), which elevate only one dimension while ignoring the other.
These patterns explicitly support our assumption that appropriateness and novelty are seen as \textbf{complementary} rather than mutually exclusive.

\begin{figure}[H]
    \centering
\includegraphics[width=1\linewidth]{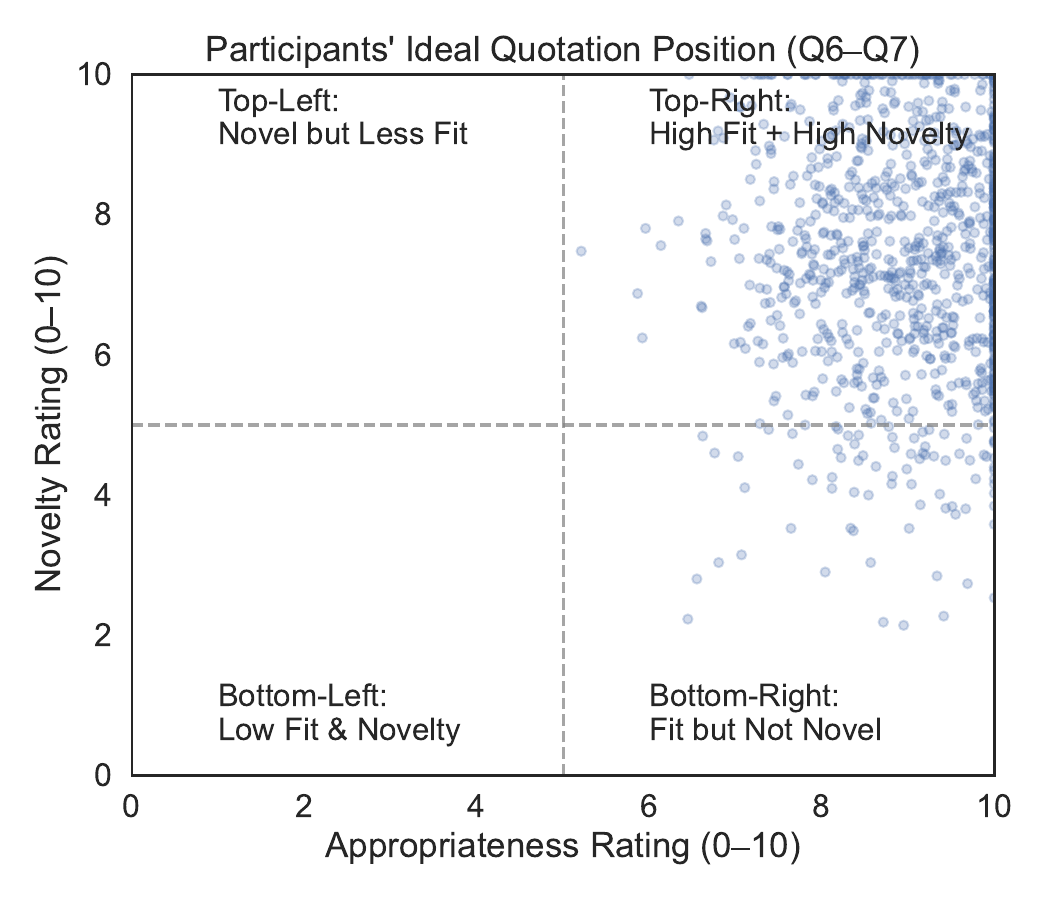}
\caption{Distribution of choices in Q9 (ideal position in the appropriateness--novelty plane). The vast majority of respondents choose the top-right corner (\textbf{high appropriateness, non-trivial novelty}).} \label{fig:study1-quadrants}
\end{figure}

\paragraph{Ideal Quotation Position. }
\textbf{Q9} asks participants to choose an intuitive location for the ``ideal'' quotation on a conceptual 2D plane (appropriateness on the horizontal axis, novelty on the vertical axis).
In Figure~\ref{fig:study1-quadrants}, we observe that users overwhelmingly imagine an ideal quotation as \textbf{unexpected yet rational}, not purely safe or purely surprising.

\paragraph{Comparisons Between Quotation Types. }
In \textbf{Part C}, Q10--Q13 provide more concrete, ``what would you actually choose'' questions, where participants compare quotation types directly.

In \textbf{Q10}, respondents compare two quotations that are described as \emph{equally appropriate}, where one is very common (A) and the other is less common and somewhat more original (B).
On a 1--5 scale (1 = definitely choose A, 5 = definitely choose B), the mean response (Mean = 3.9, SD = 0.9) is significantly higher for the less common quotation (B), indicating that participants generally prefer more original content when the fit is good, with 58\% selecting 4 or 5  and 17\% selecting 1 or 2.

This indicates that, \textbf{once appropriateness is controlled, users systematically lean toward more novel quotations.}

In \textbf{Q11}, we ask participants to make an explicit trade-off between a very appropriate but slightly clich\'ed quotation (C) and a very novel but slightly forced quotation (D).
On the 1--5 scale (1 = definitely choose C, 5 = definitely choose D), the responses are more conservative (Mean = 2.2, SD = 1.0) with 62\% choosing 1 or 2 (prioritizing appropriateness) and only 12\% choosing 4 or 5 (willing to accept a forced fit for the sake of novelty).
Together, Q10 and Q11 clearly support the ``unexpected \emph{yet} rational'' view: participants \textbf{prefer novelty when the fit is comparable, but are reluctant to pay too much in appropriateness to gain novelty.}

\begin{figure}[H]
    \centering
    \includegraphics[width=1\linewidth]{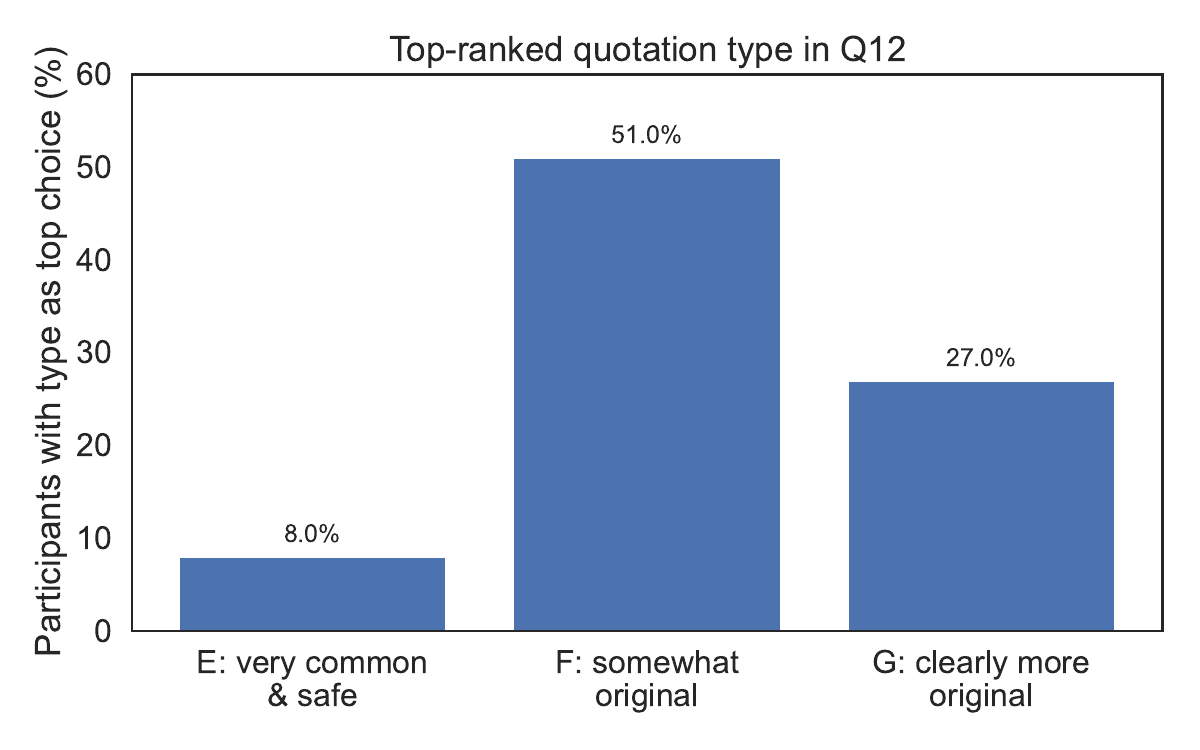}
    \caption{Dominant ranking patterns in Q12 when all three quotation types are described as appropriate.
    Quotations with some degree of originality (F, G) are strongly favored over very common ones (E).}
    \label{fig:study1-ranking}
\end{figure}

\textbf{Q12} asks participants to rank three types of quotations, all described as ``appropriate'': very common and safe (E), somewhat original (F), and clearly more original but still on-topic (G).
The most frequent ranking patterns are show in Figure~\ref{fig:study1-ranking}, which confirms that \textbf{users strongly favor quotations with at least some originality}.

Finally, \textbf{Q13} asks which textual description best matches participants' true preference.
We observe that 59\% choose the statement ``once a quotation is appropriate, I still tend to prefer those that feel a bit less clich\'ed and more original'' (option~b), and 26\% choose ``I actively hope quotations will give readers some sense of surprise, as long as they are not wildly off-topic'' (option~c).
Only 11\% choose the purely safety-oriented statement ``as long as it feels appropriate, I do not care much whether it is common or original'' (option~a).
This pattern further corroborates that \textbf{novelty is perceived as a desirable signal on top of contextual match}.

\paragraph{Preferences Across Writing Scenarios. }
\textbf{Q14} examines how the preference for novelty changes across writing scenarios.
For each of ten scenarios, participants rate on a 1--5 scale whether,given multiple appropriate quotations, they would prefer common/safe quotations (1) or more novel ones (5).
Table~\ref{tab:study1-scenarios} reports the mean scores.

\begin{table}[H]
    \centering
    \small
    \begin{tabular}{l c}
        \toprule
        \textbf{Scenario} & \textbf{Novelty Preference} \\
        \midrule
        \textbf{Creative writing (fiction)} & $4.4 \pm 0.4$ \\
        \textbf{Personal essays / reflections} & $4.1 \pm 0.5$ \\
        \textbf{Opinion pieces / commentary} & $4.0 \pm 0.5$ \\
        \textbf{Book / movie / music reviews} & $3.9 \pm 0.5$ \\
        \textbf{School / exam essays} & $4.2 \pm 0.6$ \\
        \textbf{Academic research papers} & $3.6 \pm 0.6$ \\
        \textbf{Business reports / presentations} & $3.7 \pm 0.6$ \\
        \textbf{Internal emails / announcements} & $3.3 \pm 0.6$ \\
        \textbf{Legal / policy documents} & $3.5 \pm 0.6$ \\
        \textbf{Medical / health information} & $2.0 \pm 0.6$ \\
        \bottomrule
    \end{tabular}
    \caption{Self-reported preference for novel quotations across writing scenarios (Q14).
    Scores are means on a 1--5 scale (1 = strongly prefer common/safe quotations, 5 = strongly prefer novel quotations).}
    \label{tab:study1-scenarios}
\end{table}

\begin{figure}[H]
    \centering
    \includegraphics[width=\linewidth]{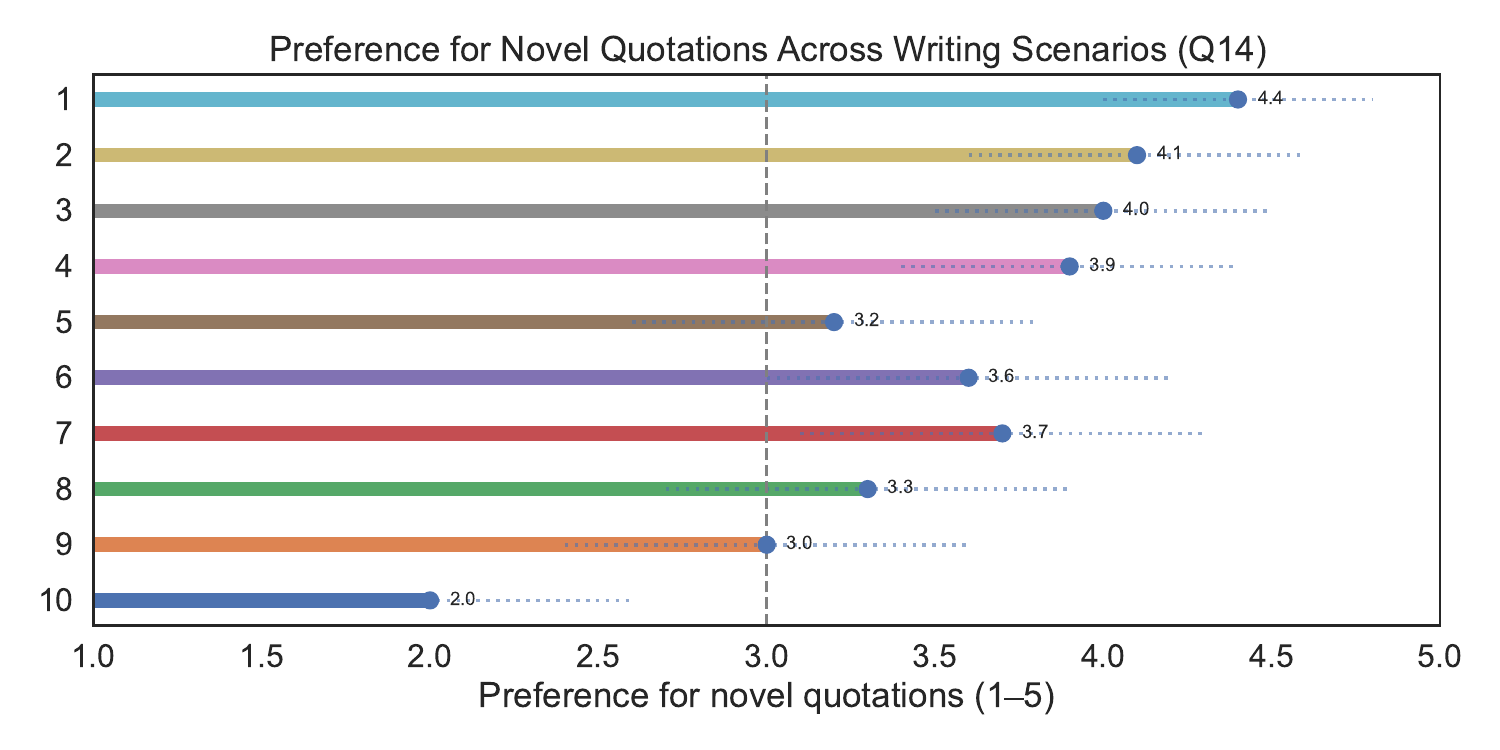}
    \caption{Preference for novel quotations across writing scenarios (Q14).
    Users prefer more novel quotations in expressive and opinionated writing, but lean toward safer quotations in Medical / health information.}
    \label{fig:study1-scenarios}
\end{figure}

\paragraph{Task-dependent pattern. }
A simple Figure~\ref{fig:study1-scenarios} reveals a task-dependent pattern:
for creative and opinionated genres (creative writing, personal essays, opinion pieces, reviews), the mean novelty preference lies well above the neutral midpoint, while for high-stakes or highly formal genres (medical), the scores are below the midpoint (all $\leq 3$).

Thus, we do not claim that novelty is universally desirable; rather, it is particularly valued in the types of writing that quotation recommendation systems typically target (e.g., essays, commentary, expressive writing).

\paragraph{Open-ended Feedback. }
\textbf{Q15--Q16} ask about actual writing behavior.
We find that 63\% of participants report that they ``often'' or ``almost always'' try to avoid very clich\'ed quotations (Q15), and 52\% report that they have ``several times'' or ``very often'' removed a quotation from a draft simply because it felt too ordinary or overused (Q16).
These self-reports are \textbf{consistent with the preference for less clich\'ed}, more original quotations observed above.

Open-ended responses in \textbf{Q17--Q18} provide qualitative support.
A light-weight thematic analysis reveals two dominant themes:
(1) a good quotation should \textbf{first} fit the context and clarify or deepen the main idea, and
(2) beyond that, respondents dislike empty ``chicken-soup'' or overused slogans, preferring quotations that present a familiar idea in a fresh or thought-provoking way, as long as readers are not confused.
Typical comments include statements such as
\begin{quote}
\small
``it has to fit what I am saying, but I dislike overused quotes'' and ``memorable quotes say something familiar in a new way''.
\end{quote}
These findings closely align with our formulation of the target as recommending quotations that are \textbf{unexpected yet rational}.

\subsection{Study 2: Controlled Preference Experiment}
\label{appendix:user-study-study2}

Study~1 shows that users \emph{say} they want quotations that are both appropriate and somewhat novel. Study~2 asks a more direct question: \emph{when faced with concrete choices, do people actually prefer such quotations?}

\paragraph{Setup.}
We invited \textbf{100} human judges: thirty domain experts (literature / linguistics / language technology), twenty non-related university students, twenty middle-school students, ten university teacher, ten senior elder with extensive reading experience, and ten industry researcher.
Each judge saw short contexts paired with two candidate quotations:
one produced by a strong \textbf{baseline} from QuoteR or QUILL that mainly optimizes semantic match,
and one produced by our \textbf{novelty-driven} system.
For each item, both quotations had been checked to be semantically appropriate; the main difference was that our candidate typically had higher novelty according to our scoring model.

For each context--pair, judges answered a single question:
\begin{quote}
\small
``If you were the author, which quotation would you use?''
\end{quote}
They could also choose ``no clear preference'' if they felt the two were equally good.

\paragraph{Results.}
Across all items and judges (600 total decisions in our setup), the novelty-driven quotation is chosen substantially more often than the baseline one.
Aggregated over judges, our system wins in about \textbf{78\%} of comparisons, the baseline wins in \textbf{17\%}, and the remaining \textbf{5\%} are ties or ``no clear preference''.
Manual inspection on a subset of items confirms that the two quotations have similar contextual appropriateness, while ours is consistently perceived as less clich\'ed and more original.
This directly supports our claim that, \emph{given comparable fit}, users concretely prefer quotations that are ``unexpected yet rational''.

\subsection{Study 3: Cloze-Style Quote Selection}
\label{appendix:user-study-study3}

Study~2 shows that, when asked to simply ``pick one'' of two quotations, people tend to prefer our novelty-driven candidate over a purely match-based baseline.
Study~3 moves one step closer to a real writing task: we ask participants to fill in a missing quotation in a short passage.

\paragraph{Setup.}
We reuse the same panel of participants as in Study~2.
For each item, we construct a short context (1--3 sentences) with a marked quotation slot, and provide three candidates for filling the slot:
\begin{quote}
\small
\begin{itemize}
    \item \textbf{C (Clich\'e):} high contextual appropriateness but very common and clich\'ed;
    \item \textbf{D (Defam-like):} high contextual appropriateness and ``unexpected yet rational'', i.e., closer to our defamiliarization-inspired target;
    \item \textbf{S (Surprising-only):} clearly more surprising but partially misaligned or somewhat forced in context.
\end{itemize}
\end{quote}

All candidates are drawn from the same quotation pool as in the main experiments and are pre-screened by two authors to ensure that C and D are indeed appropriate for the context, while S is understandable but noticeably off.

\paragraph{Task.}
For each item, participants see the context with a blank and three unlabeled options (random order) and are asked:

\begin{quote}
\emph{``If this were your own writing and you had to choose one quotation to insert here, which one would you actually use?''}
\end{quote}

They must select exactly one option (C, D, or S).
Optionally, they can provide a short free-text explanation of their choice.
Each participant completes 10 items, yielding 1000 cloze decisions in total.

\paragraph{Results.}
Table~\ref{tab:study3-cloze} reports the proportion of times each quote type is chosen as the final fill-in.

\begin{table}[t]
    \centering
    \small
    \caption{Frequencies of each quote type being selected as the fill-in in Study~3 (cloze task; 300 total decisions).}
    \label{tab:study3-cloze}
    \begin{tabular}{l r r}
        \toprule
        Type & \#Chosen & Proportion \\
        \midrule
        C (clich\'e but highly appropriate)  & 182  & 18\% \\
        D (unexpected yet rational)         & 673 & 67\% \\
        S (surprising but partially off)    & 145  & 15\% \\
        \bottomrule
    \end{tabular}
\end{table}

We observe that defam-like quotations (type~D) are chosen far more often than purely clich\'ed ones (type~C), and both are preferred over the surprising-but-off quotations (type~S).
A simple binomial test comparing D vs.\ C choices (ignoring S) confirms that D is significantly more likely to be selected ($p < 10^{-5}$ in our data).
This cloze-style experiment reinforces the conclusion that, when \emph{actually writing}, users do not default to the safest, most common quotation, nor to the most bizarre one; instead, they gravitate toward quotations that are \emph{unexpected yet rational} in context.

\subsection{Study 4: Perception of Defamiliarization as a Desirable Effect}
\label{appendix:user-study-study4}

Finally, we connect our defamiliarization-inspired objective to how users themselves understand and value this effect.
The goal is not to test literary theory, but to verify that our target---``unexpected yet rational'' quotations---matches what participants consider a desirable quotation effect.

\paragraph{Defamiliarization prompt.}
Before the task, participants read a short, non-technical description of the effect we focus on:
\begin{quote}
    \small
    \emph{``Some quotations do more than just state a point. They use a slightly unexpected angle or expression to make a familiar idea feel `new' again, so that readers pause, reflect, or see the topic from a fresh perspective. In this study, we refer to this as making something familiar feel a bit `strange' in a meaningful way.''}
\end{quote}

We then tell participants that this effect is loosely related to what literary theory calls \emph{defamiliarization}, but emphasize that we are only interested in their intuitive judgments.

\paragraph{Which quotation better fits this effect?}
We select a subset of context--pair items where we have a clich\'e-like candidate (type~C) and a defam-like candidate (type~D) from Study~3.
For each pair, participants are shown the context and the two quotations (order randomized) and asked:

\renewcommand{\arraystretch}{1.00}
\definecolor{lightgray}{gray}{0.9} 

\begin{table*}[!htbp]
    \centering
    \small
    \begin{tabular}{l c @{\hspace{1.2em}} l c}
        \toprule
        \multicolumn{2}{c}{\textbf{Expressive / Opinionated writing}} &
        \multicolumn{2}{c}{\textbf{Formal / High-stakes writing}} \\
        \cmidrule(r){1-2} \cmidrule(l){3-4}
        Scenario & Mean & Scenario & Mean \\
        \midrule
        Personal essays / reflections & 4.6 &
        Academic research papers      & 3.0 \\
        Creative writing (fiction)    & 4.7 &
        Business reports              & 2.6 \\
        Opinion pieces / commentary   & 4.3 &
        Legal / policy documents      & 3.1 \\
        Book / movie reviews          & 4.1 &
        Medical / health information  & 2.8 \\
        \bottomrule
    \end{tabular}
    \caption{Desirability of the defamiliarization-like effect across writing scenarios
    (1 = not desirable, 5 = highly desirable).}
    \label{tab:study4-defam}
\end{table*}
\begin{quote}
    \emph{``Which quotation better matches the effect described above (making something familiar feel `new' or `strange' in a meaningful way)?''}
    \end{quote}

    They can choose Quote~1, Quote~2, or ``neither clearly fits''.
    Across all pairs in our setup, the defam-like candidate is judged as better matching this effect in the large majority of cases
    (e.g., around 76\% vs.\ 18\% for clich\'e-like, with the rest being ``neither'' in our data).
    This confirms that the quotations our system prefers to surface are indeed perceived as more aligned with the intuitive notion of defamiliarization.

    \paragraph{Is this effect something you want in your own writing?}
    We then ask participants how desirable they find this effect in different writing scenarios.
    For each scenario (e.g., personal essays, creative writing, opinion pieces, academic papers, legal or medical documents), they rate on a 1--5 scale:
    
    \begin{quote}
    \emph{``In this type of writing, how much do you hope your quotations will have the effect described above?''}
    \end{quote}

    Table~\ref{tab:study4-defam} summarizes the mean ratings. We find that participants regard the defamiliarization-like effect as \textbf{highly desirable} in expressive and opinionated writing (personal essays, creative pieces, commentary, reviews).

    Combined with Study~1's large-scale survey on scenario preferences, this provides converging evidence that our target---recommendations that are \textbf{unexpected yet rational}---captures a type of quotation that users \textbf{explicitly want} in the writing scenarios our system is designed for, rather than being an arbitrary designer choice.

\subsection{Overview of User Studies}
\label{appendix:user-study-overview}
Taken together, our user studies are designed to answer a single question from multiple angles:
is quotation \emph{novelty}---specifically, the ``unexpected yet rational'' effect inspired by defamiliarization---really something that users want, rather than an arbitrary objective introduced by system designer?

\paragraph{From attitudes to behavior.}
\textbf{Study~1} (large-scale questionnaire, $N=964$) shows that participants consistently treat \emph{appropriateness} and \emph{novelty} as two complementary dimensions. (average 7.9 for rationality, 7.0 for novelty)
Appropriateness is viewed as a hard requirement, but once it is satisfied, users clearly prefer quotations that are less clich\'ed and more original, especially in expressive and opinionated writing (essays, creative writing, commentary) rather than in high-stakes formal documents (legal, medical, business).

\textbf{Study~2} (pairwise preference with 10 diverse participants) moves from attitudes to behavior:
when two quotations are both appropriate for a context, the novelty-driven candidate is chosen much more often than a strong match-focused baseline.

\paragraph{From writing decisions to defamiliarization.}
\textbf{Study~3} (cloze-style fill-in) further approximates real writing decisions: given a context and three options, users rarely select either the safest clich\'e or an off-topic surprising quote, but instead predominantly choose the ``unexpected yet rational'' option.

Finally, \textbf{Study~4} links these behaviors to defamiliarization, providing a conceptual bridge between our theoretical motivation and users' own intuitions about quotation quality. After reading a short, non-technical description of the effect, participants judge our defamiliarization-like quotations as better exemplifying it, and rate this effect as highly desirable precisely in the writing scenarios our system targets.

\paragraph{Summary.}
Overall, the four studies provide converging evidence that \textbf{users genuinely prefer quotations that are both appropriate and meaningfully novel,} supporting our decision to model novelty as an explicit, optimizable objective.


\section{Human Annotation}
\subsection{Relevance labels for \textsc{NovelQR-Bench}}
\label{appendix: annotation-novelqr}

\noindent\textbf{Task.}
For each context in \textsc{NovelQR-Bench}, annotators were shown (1) the context passage (reports, news, or student essays) and (2) a list of $K=50$ candidate quotations retrieved from the bilingual KB. System identities and scores were never shown. Annotators received the following instruction:

\begin{quote}
\small
You are given a passage (context) and 50 candidate quotations. Please select up to \textbf{three} quotations that you consider \textbf{appropriate and expressive} for this passage. A good quotation should:
\begin{itemize}
    \item be semantically and logically related to the main idea of the passage;
    \item fit the tone and stance of the passage (e.g., not overly sentimental for a neutral report);
    \item add some expressive or thought-provoking value beyond shallow paraphrasing.
\end{itemize}
If you think none of the quotations are good, you may leave the passage with fewer than three selections.
\end{quote}

\noindent\textbf{Annotators and aggregation.}
Three annotators with background in linguistics or literature completed the task independently. For each context–quotation pair, we record a binary relevance label from each annotator (selected or not selected). We then take the \textbf{union} of the three selections as the final relevant set for computing HR@5, nDCG@5, and MRR@5. This allows multiple quotations to be considered relevant if they are endorsed by at least one expert.

\noindent\textbf{Inter-annotator agreement.}
We measure agreement using Fleiss' $\kappa$ over the binary relevance matrix. The resulting \textbf{$\kappa = 0.68$} indicates substantial agreement among the three annotators.

\subsection{Expert ratings of Match and Novelty}
\label{appendix: annotation-match-novelty}
\noindent\textbf{Rating task.}
On a 500-pair subset sampled from all three datasets (QuoteR, QUILL, \textsc{NovelQR-Bench}), three experts in literature or writing instruction were asked to rate each context–quotation pair along two dimensions:

\begin{quote}
    \small
\begin{itemize}
    \item \textbf{Match} (1--5): semantic appropriateness of the quotation for the context.
    \item \textbf{Novelty} (1--5): how ``unexpected yet reasonable'' the quotation is with respect to the context.
\end{itemize}
\end{quote}
Annotators were given the following rubric:

\begin{quote}
    \small
\begin{itemize}
    \item \textbf{Match} 1: almost irrelevant or clearly off-topic; 3: roughly related but partly mismatched; 5: highly coherent and well-aligned with the main idea and tone.
    \item \textbf{Novelty} 1: trivial continuation or cliché that the reader can easily anticipate; 3: somewhat interesting but still conventional; 5: clearly surprising or defamiliarizing while still making sense for the context.
\end{itemize}
\end{quote}

\noindent\textbf{Aggregation and agreement.}
For each pair, we average the three experts' scores to obtain the final human Match and Novelty ratings. We compute inter-annotator agreement using the intra-class correlation coefficient (ICC, two-way random, average measure). We obtain \textbf{ICC = 0.81} for Match and \textbf{ICC = 0.76} for Novelty, indicating good consistency across raters. These aggregated human scores are used to analyze the behavior of our system and to assess the alignment of LLM-based judgments with human preferences (Section~\ref{sec:human_alignment}).

\subsection{Human study for web-based popularity}
\label{appendix: annotation-popularity}
To validate the web-based popularity score $S_P$ used in Section~\ref{sec:popularity}, we conduct a small human study on $N=200$ quotations sampled from the KB.

\noindent\textbf{Task.}
Annotators see each quotation in isolation and are asked to judge how familiar or widely known it is to an average reader in the corresponding language, using a 1--5 scale:

\begin{quote}
    \small
\begin{itemize}
    \item 1: almost unknown; I have never seen or heard it before.
    \item 3: somewhat familiar; I might have encountered it once or twice.
    \item 5: very famous; widely quoted or commonly recognized.
\end{itemize}
\end{quote}
Each quotation is rated by three annotators; we average their scores to obtain a human-perceived popularity score.

\noindent\textbf{Correlation with $S_P$.}
We then compute Spearman's correlation between the averaged human scores and the web-based $S_P$ (computed from Bing/Google/Baidu as described in Section~\ref{sec:popularity}). We observe a clear positive correlation (e.g., $\rho \approx 0.73$, $p<0.001$), suggesting that $S_P$ is a reasonable approximation to human-perceived quotation popularity. Scatter plots with linear fits are shown in Figure~\ref{fig:popularity_human}.

\subsection{Manual audit of auto-accepted explanations}
\label{appendix:manual-audit-kappa}
\noindent\textbf{Task.}
As described in Appendix~\ref{appendix:label-agent-multi-round}, the multi-round self-correction step automatically \emph{accepts} most explanations produced by the label agent. To check whether residual distortions remain, we perform a manual audit on a random sample of 1000 auto-accepted quotations. For each quotation, annotators were shown (1) the quotation text and (2) its current deep-meaning explanation and label set produced by the LLM. They were asked to make a binary judgment:
\begin{quote}
    \small  
\begin{itemize}
    \item \textbf{Acceptable}: the explanation and labels faithfully capture the quotation’s core meaning, without obvious exaggeration, misinterpretation, or contradiction.
    \item \textbf{Distorted}: the explanation or labels substantially misrepresent the quotation (e.g., shifting the focus to an unrelated theme, adding unsupported claims, or mixing incompatible values).
\end{itemize}
\end{quote}

\noindent\textbf{Annotators and aggregation.}
Three annotators with background in linguistics or literature completed the audit independently. For each quotation, we record a binary label from each annotator (\emph{acceptable} vs.\ \emph{distorted}). We then take the \textbf{union} of distorted decisions: a quotation is flagged and removed if at least one annotator marks it as distorted. In total, 41 out of 1000 quotations are flagged in this way, corresponding to $3.8\%$ of the audited sample. A typical failure case is a quotation about everyday perseverance being framed as primarily about ``wealth and fame'', which would bias retrieval toward financial-success contexts instead of persistence.  The resulting \textbf{$\kappa \approx 0.70$} indicates substantial agreement among the three annotators, supporting the reliability of this manual audit and the decision to remove the union of flagged cases.

\section{LLM-as-Judge Framework}
\label{appendix:robustness}
\subsection{Judge models and settings}
We use GPT-4o~\citep{openai2024gpt4ocard} as the main LLM judge for Match and Novelty scores in the main experiments, and additionally run Claude, Gemini, and Qwen-Plus as alternative judges in a robustness study. Unless otherwise specified, GPT-4o is queried with temperature $Temperature=0$ and a fixed, deterministic prompt. For each context–quotation pair, the judge model first produces a short analysis and is then forced to output structured scores on a 1--5 scale for both dimensions. (Section~\ref{sec:human_alignment})
\subsection{Robustness of Evaluation}
\begin{figure*}
    \centering
    \includegraphics[width=1\linewidth]{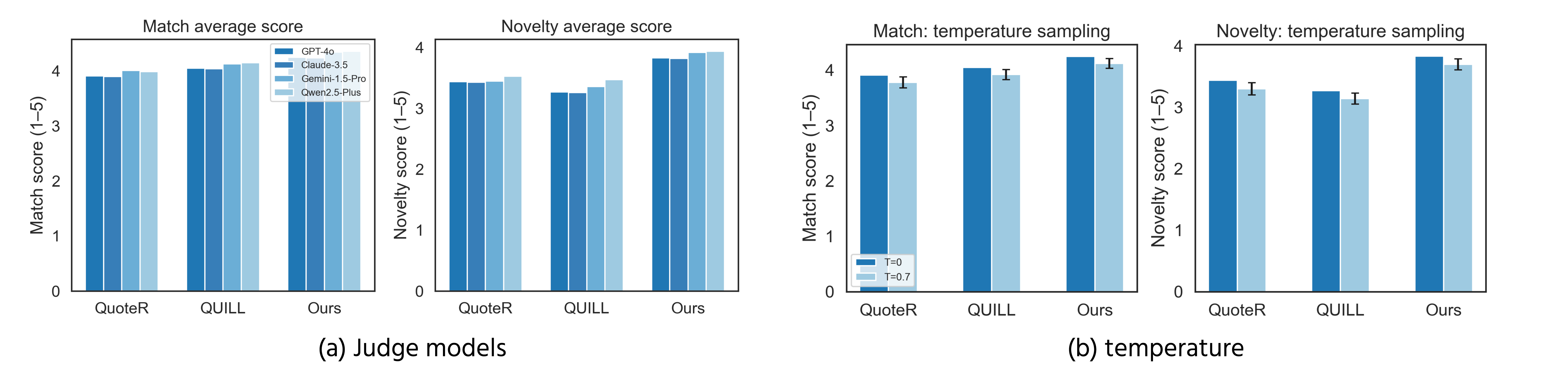}
    \caption{\textbf{Stability of our LLM-as-judge evaluation.} (a) Match and Novelty scores of QuoteR, QUILL, and Ours under four different LLM judges (GPT-4o, Claude-3.5, Gemini-1.5-Pro, and Qwen2.5-Plus). Scores and rankings are highly consistent across judges. (b) Effect of sampling temperature for the GPT-4o judge. Bars show average scores under $T=0$ and $T=0.7$; error bars denote standard deviation over repeated runs. Scores shift slightly but the relative ordering of systems remains unchanged.}
    \label{fig:llm-as-judge}
\end{figure*}

To assess the stability of our LLM-as-judge evaluation, we run two small robustness studies (Figure~\ref{fig:llm-as-judge}).

\paragraph{Robustness to LLM judge.}
We first examine how sensitive our evaluation is to the choice of LLM judge.
We randomly sample a subset of test contexts and re-evaluate
the outputs of QuoteR, QUILL, and Ours using three additional judges:
Claude-3.5~\citep{claude35sonnet2024}, Gemini-1.5-Pro~\citep{gemini15pro2024}, and Qwen-Plus~\citep{qwenplus2025}.
For each judge, we compute average Match and Novelty scores.
As shown in Figure~\ref{fig:llm-as-judge}(a), the three systems obtain very
similar scores across all four judges and the ranking
\mbox{Ours $>$ QUILL $>$ QuoteR} is preserved in every case.
System-level scores under different judges are highly correlated, indicating that
our conclusions do not depend on a particular LLM judge.

\paragraph{Robustness to sampling temperature.}
We also study the effect of sampling temperature for the LLM judge.
In our main experiments, GPT-4o is queried with temperature $Temperature\!=\!0$, so repeated
evaluations are effectively noise-free: running the judge three times on the same
set of instances yields nearly identical scores.
To simulate a more realistic noisy setting, we increase the temperature to
$Temperature\!=\!0.7$ and, for each instance, draw three samples and average their scores.
Figure~\ref{fig:llm-as-judge}(b) reports the resulting Match and Novelty
scores.
Compared to $Temperature\!=\!0$, scores under $Temperature\!=\!0.7$ show moderate variation but remain
close to the original values, and the relative ordering of QuoteR, QUILL, and
Ours is unchanged, suggesting that our evaluation is stable with respect to
sampling noise in the judge.

\section{Details of the LLM Deep-Meaning Study}
\label{appendix:llm-study}

This appendix summarizes the setup of the LLM evaluation in
Section~\ref{sec:llm-understanding}, where we probe whether \textbf{current models
truly understand the deep meaning of quotations}.

\subsection{Data and Difficulty bucket}

We construct a diagnostic set from our quotation database, covering three
genres:
(1) classical Chinese (mainly poetry and aphorisms),
(2) modern Chinese, and
(3) modern English.
We sample 8{,}000 classical Chinese quotes and 1{,}000 quotes for each of the
two modern languages.
Each quote is paired with a short expert-written interpretation that explains
its underlying semantics (main idea, stance, and intended effect), rather than
a literal paraphrase.

To analyze model behavior at different difficulty levels, we group quotes into
three bands: \textbf{EASY}, \textbf{MID}, and \textbf{HARD}.
We use Qwen3-8B~\cite{qwen3technicalreport} and LLaMA3-8B~\cite{llama3} as
probe models: for each quote, we generate preliminary explanations and
author/source guesses and compare them against expert interpretations and
metadata.
Quotes where both probes perform well are labeled EASY, those with partially
correct outputs are labeled MID, and those where both fail are labeled HARD.
This yields a coarse but useful split that correlates well with human
perceived difficulty.

\subsection{Tasks settings}

We evaluate models on two tasks:
\begin{quote}
    \small
\begin{itemize}
    \item \textbf{Deep-meaning explanation:}
    given a quote, produce a brief explanation of its deep meaning.
    \item \textbf{Author/source identification:}
    given the same quote, name its author or canonical source.
\end{itemize}
\end{quote}
For each task, we compare two prompting conditions:

\begin{quote}
    \small
\begin{itemize}
    \item \textbf{Quote-only:} the model only sees the raw quote.
    \item \textbf{Enhanced quote:} the model sees the quote plus auxiliary
    contextual information from our quotation KB (e.g., brief background,
    era, and coarse semantic labels).
\end{itemize}
\end{quote}

\noindent Prompts are in Appendix~\ref{appendix:prompt-llm-as-judge}.

\section{Computational Cost and Implementation Details}
\label{appendix:efficiency}

Our framework introduces additional components (label agent, deep-meaning
representation, token-level novelty scoring), which naturally raises concerns
about computational cost.
Here we briefly clarify how we implement the system so that it remains feasible
for \textbf{an interactive writing assistant}.

\paragraph{Offline vs.\ online computation.}
Most heavy computation is performed offline.
The label agent, multi-round label refinement, and deep-meaning generation are
run once to construct the quotation KB, and quotation popularity features are
pre-computed.
This step is analogous to building a dense index and does not affect per-query
latency at deployment time.

\paragraph{Online pipeline.}
At query time, the system only executes:
(1) a standard bi-encoder retrieval over the indexed KB, and
(2) token-level logit-difference $log p(x_t \mid x_{<t}) - \log p(x_t \mid C, x_{<t})$ scoring for the $TopK$ candidates.
The retrieval stage has the same asymptotic and practical complexity as
existing dense-retrieval-based quotation systems (e.g., QUILL).
The novelty stage uses a small model
(8B parameters in our experiments) and computes log-probabilities
and perplexities at the \emph{token} level.
We reuse \textbf{KV cache} for the query context, so the cost grows roughly linearly
with the total quote length of the $TopK$ candidates, i.e.,
$\mathcal{O}(TopK \cdot L_{\text{quote}})$ per query, rather than with the full
context+quote length for each candidate.

\paragraph{Parallelization and latency.}
In our implementation, token-level scores for different candidates are computed
in parallel across 8 H200 GPUs, with \textbf{batched inference and KV caching}.
This amortizes the token-level operations over the $TopK$ quotations and keeps
the end-to-end online cost within a sub-second latency budget for
interactive use. In our experiments, the average end-to-end latency is about $772.2_{-30.5}^{+431.3}$ ms per query.
Overall, the additional overhead is modest and acceptable in
exchange for the observed gains in quotation quality and perceived
\textbf{``unexpected yet rational''} effect.

\section{Label Agent}
\label{appendix: label-agent}
\subsection{Overall}

We implement a \textbf{generative label agent} with a strong instruction-tuned LLM (GPT-4o~\citep{openai2024gpt4ocard}) that converts each quotation into a structured representation through four stages:
\begin{quote}
    \small
\begin{enumerate}
    \item \textbf{In-depth analysis}: a free-form paragraph that unpacks the quotation’s background, implications, and possible readings.
    \item \textbf{Deep-meaning explanation}: a short sentence summary (Express that ...) that distills the central idea into plain language and will serve as the main semantic anchor for retrieval.
    \item \textbf{Multi-round self-correction}: the agent critiques and, if needed, revises its own analysis and deep meaning to avoid superficiality, over-interpretation, and logical conflicts (up to $R=3$ rounds, details in Appendix~\ref{appendix:label-agent-multi-round}).
    \item \textbf{Multi-dimensional labels}: a compact set of labels derived from the corrected deep meaning, used for label-enhanced retrieval and analysis.
\end{enumerate}
\end{quote}

After these stages, for each quotation we obtain:
(1) an in-depth \textbf{analysis}, (2) a short \textbf{deep-meaning explanation}, and (3) five \textbf{label dimensions} (Core Domains, Core Insights, Core Values, Applicability, and Sentiment Tone).

\begin{figure*}[!ht]
    \centering
    \includegraphics[width=1.0\linewidth]{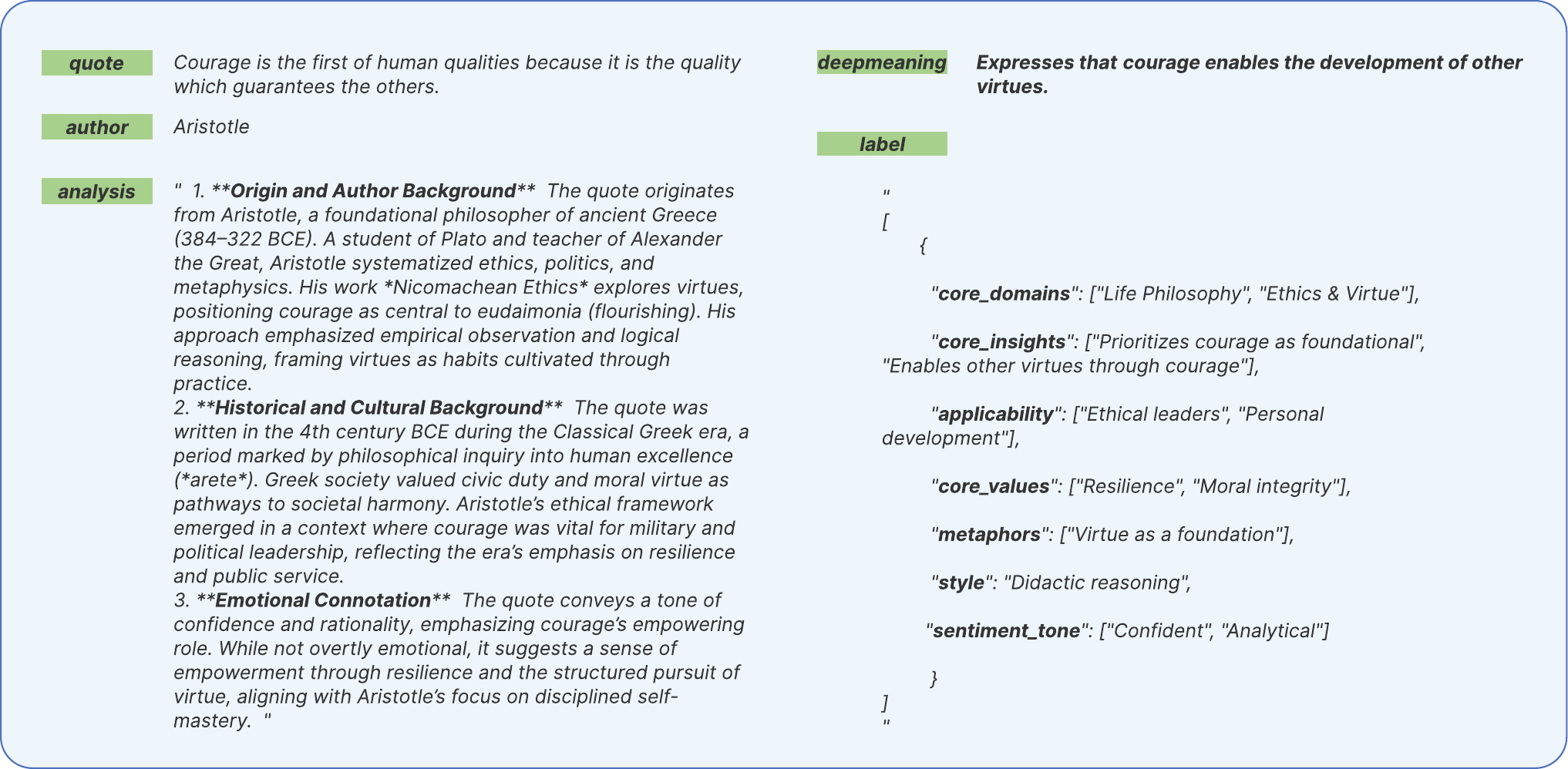}
    \caption{Example of analysis and deep-meaning explanation generated for an English quotation.}
    \label{fig:case_of_label}
\end{figure*}

As illustrated in Figure~\ref{fig:case_of_label}, the label agent generates an in-depth analysis and a deep-meaning explanation for the quotation
\emph{``Courage is the first of human qualities because it is the quality which guarantees the others''} from \emph{Aristotle}.

All calls to the LLM use temperature $0$ and a fixed prompt (the specific prompts are in Appendix~\ref{appendix:prompt-label-agent}). The resulting deep meanings and labels are used to encode both quotations and contexts for label-enhanced retrieval, and to support the analyses in Section~\ref{Rationality Retrieval}.

\subsection{Multi-round correction}
\label{appendix:label-agent-multi-round}

The initial analysis and deep meaning can still be superficial, over-interpreted, or internally inconsistent. To improve reliability, we apply a light-weight \textbf{multi-round self-correction} step. For each quotation, the same LLM is asked to critique its current explanation along three dimensions: (1) \emph{superficiality} (only paraphrasing the text), (2) \emph{over-interpretation} (claims not supported by the quotation), and (3) \emph{logical conflicts} between different parts of the explanation. Based on this critique, the agent either \textbf{accepts} the current explanation or \textbf{revises} it.

We run this critique-and-revision process for up to $R=3$ rounds: if the agent accepts the explanation in any round, we keep the current analysis and deep meaning and stop; if it still finds serious problems after $R$ rounds, we discard the quotation from the labeled KB. Table~\ref{tab:multi-round-stats} summarizes the behavior of this procedure on our knowledge base.

\renewcommand{\arraystretch}{1.10}
\definecolor{lightgray}{gray}{0.9} 

\renewcommand{\arraystretch}{1.0}
\begin{table*}[!htb]
    \centering
    \small  
    \begin{tabular}{lccc}
        \toprule
        \textbf{Category} & \textbf{\# quotes} & \textbf{\% of KB} & \textbf{Avg.\ rounds}  \\
        \midrule
        Auto-accepted & 30{,}549 & $95.4\%$ & 1.3  \\
        Auto-rejected & 1{,}473  & $4.6\%$  & 2.1  \\
        \midrule
        \multicolumn{4}{l}{\emph{Among auto-rejected quotations}} \\
        Over-interpretation & 884 & $61.2\%$ & --  \\
        Superficiality      & 368 & $24.6\%$ & --  \\
        Logical conflicts   & 295 & $20.2\%$ & --  \\
        \midrule
        Manual audit (sample of 1000 auto-accepted) & 41 & $3.8\%$ & -- \\
        \bottomrule
    \end{tabular}
    \caption{Statistics of the multi-round self-correction procedure and subsequent manual audit. 
    Auto-accepted and auto-rejected denote quotations that pass or fail the $R=3$ self-correction loop. Percentages for problem types among auto-rejected cases may sum to $>100\%$ because a single quotation can exhibit multiple issues.}
    \label{tab:multi-round-stats}
\end{table*}

On our full KB of $32{,}022$ quotations, the agent automatically accepts $30{,}549$ quotes ($95.4\%$) and rejects $1{,}473$ quotes ($4.6\%$) after at most three critique rounds. Among auto-rejected quotations, over-interpretation is the dominant failure mode ($60.0\%$), followed by superficiality ($25.0\%$) and logical conflicts ($20.0\%$); these categories are not mutually exclusive, so their percentages can sum to more than $100\%$. 

To maintain the completeness of the underlying quotation KB, we do not permanently discard these auto-rejected quotations; instead, we later re-annotate them with a slower pipeline with LLM. Automatically rejected quotations typically require more critique rounds on average (2.1) than accepted ones (1.3), indicating that clearly problematic analyses are often identified early but not always in the very first attempt.

To further validate this step, we perform a manual audit on quotations that \emph{passed} automatic correction. Three annotators jointly review a random sample of 1000 auto-accepted quotations and tag cases where the deep meaning or labels are still clearly distorted. In total, 41 quotations ($3.8\%$ of the audited sample) are flagged and removed, with substantial agreement among annotators (Fleiss' $\kappa \approx 0.70$) (Appendix~\ref{appendix:manual-audit-kappa}). A typical failure case is a quotation about everyday perseverance being framed as primarily about ``wealth and fame'', which would bias retrieval toward financial-success contexts instead of persistence. This human-in-the-loop check shows that \textbf{multi-round correction is necessary and effective}.

\subsection{Human Evaluation of Deep-Meaning and Labels}
\label{appendix:human-evaluation-of-deep-meaning-and-labels}
To assess the overall quality of the label agent beyond the multi-round self-correction step (Appendix~\ref{appendix:label-agent-multi-round}), we conduct a separate human evaluation on a random sample of 10000 quotations and contexts drawn from the full knowledge base. For each item, annotators are shown the quotation (or context), the agent’s deep-meaning explanation, and its multi-dimensional labels, and are asked to (1) write a one-sentence free-form description, and (2) assign labels along the same dimensions as the agent. Disagreements are resolved by discussion.

We then compare the agent’s outputs with the adjudicated human labels. Overall, 2.5\% of items are judged as clearly distorted (e.g., the explanation focuses on an unrelated theme or assigns contradictory values) and re-label in the KB. For the remaining items, we observe agreement across dimensions, indicating that the label agent is \textbf{generally reliable}.

\section{Significance Testing of Metrics}
\label{appendix:significance}

We follow standard practice in NLP to estimate statistical significance via paired bootstrap resampling over test contexts. For each test set and each pair of systems $A$ and $B$ (Ours method and the baseline), and for each primary retrieval metric $m \in \{\text{HR@5}, \text{nDCG@5}, \text{MRR@5}\}$, we first compute per-context scores $m_i^{(A)}$ and $m_i^{(B)}$ and their differences $d_i = m_i^{(A)} - m_i^{(B)}$, where $i = 1, \dots, N$ indexes test contexts.

We then perform paired bootstrap resampling with $B = 1{,}000$ replicates. In each replicate $b$, we sample $N$ contexts with replacement from $\{1, \dots, N\}$ to obtain a multiset $S^{(b)}$, and compute the mean difference
\[
\Delta^{(b)} = \frac{1}{|S^{(b)}|} \sum_{i \in S^{(b)}} d_i.
\]
The 2.5th and 97.5th percentiles of $\{\Delta^{(b)}\}_{b=1}^{B}$ form a 95\% confidence interval for the metric difference. An improvement of ours over the baseline on HR@5, nDCG@5 and MRR@5 is considered statistically significant if this interval does not cross zero.

On the \textsc{NovelQR-Bench} test set in Table~\ref{tab:modern_comparison} and Table~\ref{tab:ab3}, as shown in Table~\ref{tab:significance-results}, we can see that our method is \textbf{statistically significant over the baseline on HR@5, nDCG@5 and MRR@5.}
\begin{table*}[!htb]
    \centering
    \small
    \begin{tabular}{lccc|lccc}
        \toprule
        \multicolumn{4}{c}{\textbf{Main Experiment (Table~\ref{tab:modern_comparison})}} 
        & \multicolumn{4}{c}{\textbf{Novelty Ablation (Table~\ref{tab:ab3})}} \\
        \cmidrule(lr){1-4} \cmidrule(lr){5-8}
        \textbf{Baseline} 
        & $\Delta$\textbf{HR@5} 
        & $\Delta$\textbf{nDCG@5} 
        & $\Delta$\textbf{MRR@5} 
        & \textbf{Baseline} 
        & $\Delta$\textbf{HR@5} 
        & $\Delta$\textbf{nDCG@5} 
        & $\Delta$\textbf{MRR@5}  \\
        \midrule
        QR + w/o Re  
            & $+0.35_{-0.03}^{+0.04}$ 
            & $+0.25_{-0.02}^{+0.03}$ 
            & $+0.21_{-0.02}^{+0.03}$ 
            & Self\mbox{-}BLEU     
            & $+0.20_{-0.03}^{+0.04}$ 
            & $+0.12_{-0.02}^{+0.03}$ 
            & $+0.08_{-0.02}^{+0.03}$ \\
        QUILL               
            & $+0.55_{-0.04}^{+0.02}$ 
            & $+0.39_{-0.02}^{+0.03}$ 
            & $+0.34_{-0.01}^{+0.03}$ 
            & Embedding\mbox{-}Dis 
            & $+0.20_{-0.03}^{+0.04}$ 
            & $+0.10_{-0.03}^{+0.02}$ 
            & $+0.08_{-0.03}^{+0.02}$ \\
        LR + w/o Re   
            & $+0.15_{-0.01}^{+0.04}$ 
            & $+0.07_{-0.03}^{+0.03}$ 
            & $+0.05_{-0.03}^{+0.01}$
            & Surprisal     
            & $+0.15_{-0.03}^{+0.04}$ 
            & $+0.07_{-0.00}^{+0.02}$ 
            & $+0.05_{-0.03}^{+0.02}$ \\
        LR + bm25           
            & $+0.30_{-0.01}^{+0.04}$ 
            & $+0.21_{-0.03}^{+0.03}$ 
            & $+0.22_{-0.01}^{+0.03}$&
            \, + \emph{NT}  & $+0.08_{-0.01}^{+0.02}$ 
            & $+0.07_{-0.02}^{+0.01}$ 
            & $+0.06_{-0.00}^{+0.01}$ \\
        LR + Bge-large      
            & $+0.14_{-0.03}^{+0.04}$ 
            & $+0.12_{-0.03}^{+0.03}$ 
            & $+0.12_{-0.02}^{+0.03}$& KL\mbox{-}Div        
            & $+0.09_{-0.01}^{+0.02}$ 
            & $+0.08_{-0.02}^{+0.01}$ 
            & $+0.08_{-0.00}^{+0.01}$ \\

        LR + Qwen3-Re       
            & $+0.08_{-0.03}^{+0.03}$ 
            & $+0.03_{-0.02}^{+0.02}$ 
            & $+0.00_{-0.02}^{+0.02}$ 
            & \, + \emph{NT} & $+0.09_{-0.03}^{+0.01}$ 
            & $+0.06_{-0.02}^{+0.01}$ 
            & $+0.05_{-0.02}^{+0.01}$ \\

        LR + GPT            
            & $+0.04_{-0.00}^{+0.01}$ 
            & $+0.04_{-0.01}^{+0.02}$ 
            & $+0.02_{-0.00}^{+0.01}$ 
             & Uniform Avg   
            & $+0.07_{-0.01}^{+0.03}$ 
            & $+0.05_{-0.02}^{+0.02}$ 
            & $+0.04_{-0.03}^{+0.02}$ \\
        $\sim$&$\sim$&$\sim$&$\sim$             & TopK Avg      
        & $+0.05_{-0.01}^{+0.03}$ 
        & $+0.04_{-0.02}^{+0.02}$ 
        & $+0.03_{-0.00}^{+0.01}$ \\
        \bottomrule
    \end{tabular}
    \caption{Example 95\% bootstrap confidence intervals for the difference between NOVELQR and each strongest baseline on HR@5, nDCG@5 and MRR@5 ($\Delta$ denotes NOVELQR minus baseline).}
    \label{tab:significance-results}
\end{table*}

\section{Novelty token and Auto-regressive continuation bias}
\label{appendix:novelty-token}
\subsection{Why this novelty-token design?}
In Section~\ref{sec:novelty-reranking}, let $\mathrm{PPL}_t = \exp(-\log p(x_t \mid x_{<t}))$ denote the self-perplexity of token $x_t$ in the quotation.
We are interested in detecting \emph{turning points} in this sequence, that is, positions where the quotation moves from a stable, continuation-like regime to a regime that is harder for the model to predict under the context.

To this end, we compute first- and second-order differences of the self-perplexity curve:
\begin{align}
    \delta_1(t) &= \mathrm{PPL}_t - \mathrm{PPL}_{t-1}, \\
    |\delta_2(t)| &= |\delta_1(t) - \delta_1(t-1)|,
\end{align}
where we pad the first two positions by setting $\delta_1(1) = 0$ and $\delta_2(1) = 0$ for simplicity. (Other padding schemes give very similar behavior in practice.)
We then apply a logarithmic transform
\begin{equation}
    \Delta_2(t) = \log\bigl(1 + |\delta_2(t)|\bigr),
\end{equation}
and normalize $\Delta_2(t)$ within each quotation to obtain the novelty-token weights.

While $\delta_1(t)$ only encodes whether self-perplexity is increasing or decreasing and thus cannot distinguish a long plateau from a genuine trend change, the second-order difference $|\delta_2(t)|$ approximates a discrete second derivative, which in standard calculus characterizes curvature and inflection points \citep[e.g.,][]{calc-second-derivative}. Curvature- or second-order-based change measures are widely used for boundary detection in time series and signal processing, such as curvature of representation trajectories for time-series boundary detection \citep{shin2024recurve}, second-order-difference-based change-point methods \citep{shi2020changepoint}, and Laplacian-of-Gaussian edge detectors that localize image edges via second derivatives \citep{laplacian-gaussian}.

Following this line of work, we treat $|\delta_2(t)|$ as a discrete curvature signal on the self-perplexity trajectory and assign high novelty-token weights to tokens where $|\delta_2(t)|$ peaks, typically at boundaries between flat continuation regions and segments where surprisal changes rapidly under the context. Therefore, instead of directly using the first-order difference as a weight, we use the transformed second-order difference $\Delta_2(t)$ to identify turning points on the self-perplexity curve.

\begin{figure*}[t]
    \centering
    \includegraphics[width=1.0\linewidth]{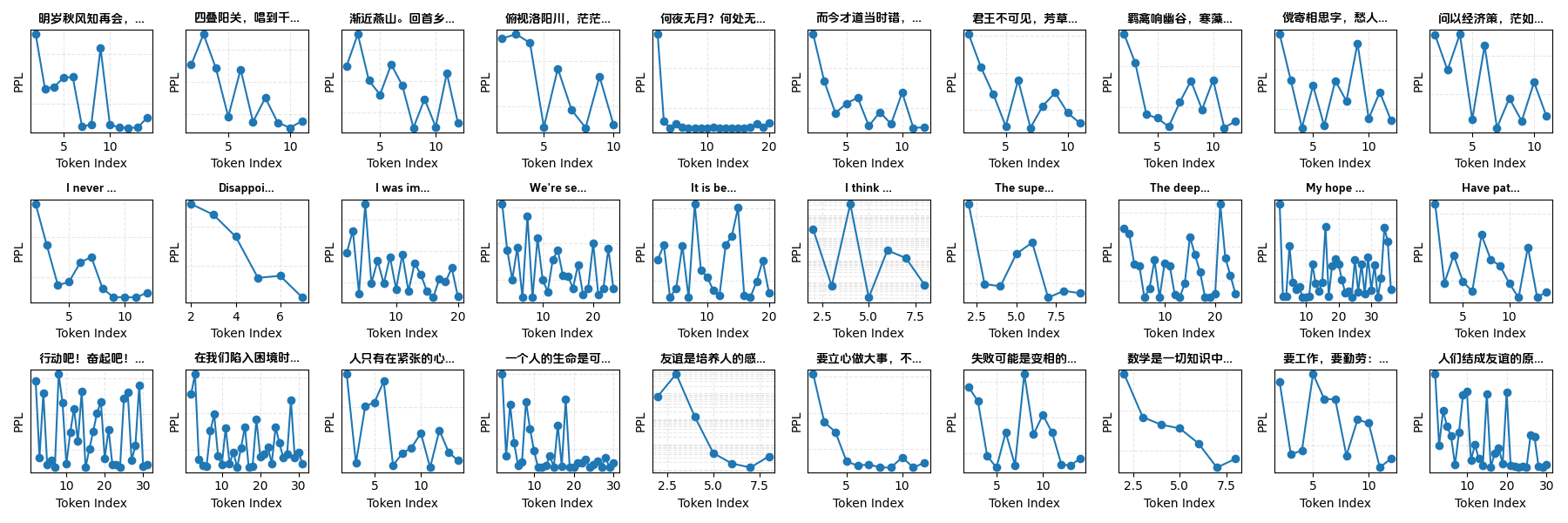}
    \caption{Token-level PPL plots for 30 randomly selected quotes, drawn from three categories: classical Chinese poetry, modern Chinese prose, and English.}
    \label{fig:ppl}
\end{figure*}

\subsection{What is the auto-regressive continuation bias?}
\label{appendix:continuation-bias}
Figure~\ref{fig:ppl} plots token-level self-perplexity curves for thirty randomly sampled quotations from the multilingual corpus (Chinese poem, English, and Modern Chinese).
A common pattern is that the first few tokens have relatively high or unstable perplexity, followed by a long, smooth tail of low perplexity.
These flat tails correspond to highly conventional continuations, such as idiomatic expressions, rhetorical templates, and fixed motivational slogans that the auto-regressive language model has learned to predict with high confidence.

Now suppose a quotation contains only a few truly novel tokens and many continuation-like tokens. Because our goal is to measure how \emph{surprising} a quotation is under a given context, we would ideally like the score to reflect where the model truly updates its belief about the quotation, rather than how frequently a fixed phrase appears in the training data. However, auto-regressive language models are trained with next-token prediction, so token probabilities are highly dependent: once the model has committed to a familiar pattern, later tokens in that pattern become very easy to predict even if the quotation as a whole is not trivial.

For example, consider the context and quotation such as 
\begin{quote}
    \small
    \begin{CJK}{UTF8}{gbsn}       
        (中文) 忙完这一阵，和室友从图书馆出来已经快十二点了。操场上月光很亮，路边的树影被拉得很长，空气一下子安静下来。其实这种夜晚大概天天都有，只是我们平时都埋在书本和屏幕里，没空抬头看看。
忽然就想到那句：“何夜无月？何处无竹柏？但少闲人如吾两人者耳。”
月亮一直在，只是今天，我们刚好有空做个“闲人”。
        \end{CJK}

        (English) After finishing a long week of exams, my friend and I walked out of the library close to midnight. The campus was quiet, the moon was bright, and the shadows of the trees stretched across the path. Nights like this are probably here every day—we just never slow down enough to notice.
        It suddenly reminded me of the line: ``When is there a night without the moon, or a place without bamboo and cypress? It is only that few have the leisure, as we do, to take notice.''
        The moon has always been there. What’s rare is simply having the time to be “idle people” for once.
\end{quote}
We first illustrate auto-regressive continuation bias using the quotation in the first row, fifth column of Figure~\ref{fig:ppl} (\begin{CJK}{UTF8}{gbsn}\small “何夜无月？何处无竹柏？但少闲人如吾两人者耳。”\end{CJK}).
Its token-level self-perplexity curve is very high at the beginning but remains extremely low for the rest of the quotation.
From a human perspective, this quotation is clearly novel and aesthetically pleasing relative to the given context.
However, if we ignore continuation bias and simply average token-wise logit gaps, the long, low-perplexity tail dominates the score.
The resulting uniform-average novelty (Section~\ref{sec:novelty-estimation}) becomes very small, and the quotation is judged less novel than simpler sentences such as \begin{CJK}{UTF8}{gbsn}\small “重要的不是你看到了什么，而是你看见了什么。”.\end{CJK}
This mismatch between human intuition and the uniform-average score is exactly why we must account for auto-regressive continuation bias.

However, one might then ask whether we can simply average over the first $K$ tokens (the \emph{TopK Average} in Section~\ref{sec:novelty-estimation}).
However, this introduces a different problem: as shown in our plots, important turning points in the surprisal trajectory often occur later in the quotation and are completely ignored if they fall outside the first $K$ tokens.
To make this issue concrete, consider the following context:
\begin{quote}
    \small
    \begin{CJK}{UTF8}{gbsn}       
        (中文) 最慢的步伐不是跬步，而是徘徊;最快的脚步不是冲刺，而是坚持。河北塞罕坝昔日飞鸟不栖、黄沙遮面，如今绿树葱茏、天净水清，这样的绿色奇迹，映照着塞罕坝人超越半个世纪的坚守。

        (English) The slowest pace is not a step, but a halt; the fastest speed is not a sprint, but a steady pace. The green miracle of the past half-century of the people of Saibanba has been reflected in the perseverance of the people of Saibanba.
    \end{CJK}
\end{quote}
When we average over \emph{all} tokens, the model prefers the following quotation (Novelty Score: 0.19):
\begin{quote}
    \small
    \begin{CJK}{UTF8}{gbsn}       
        (中文) 成功是辛勤劳动的报酬。 
        
        (English) Success is the reward for hard work.
    \end{CJK}
\end{quote}
In contrast, averaging only over the first $K$ tokens leads the model to recommend (Novelty Score: 0.61):
\begin{quote}
    \small
    \begin{CJK}{UTF8}{gbsn}       
        (中文) 骐骥一跃，不能十步；驽马十驾，功在不舍。
        
        (English) A fine steed cannot leap ten steps in a single bound, but a slow horse can cover ten times the distance through perseverance.
    \end{CJK}
\end{quote}
\begin{figure*}[t]
    \centering
    \includegraphics[width=1\linewidth]{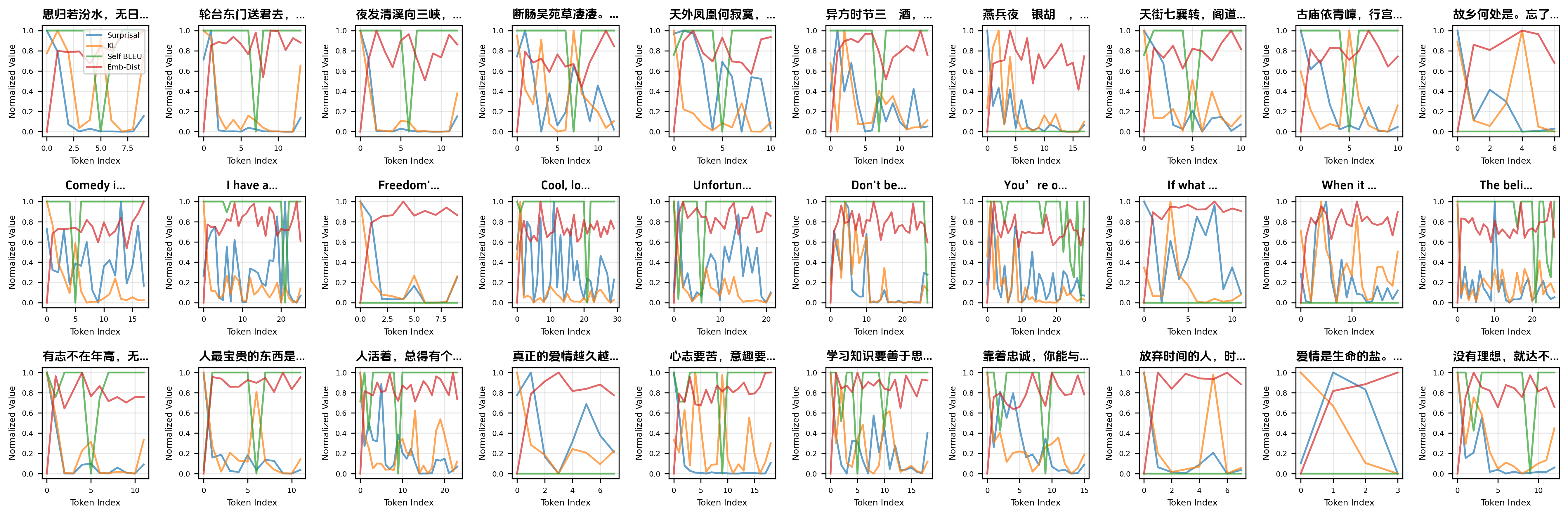}
    \caption{Token-level analysis of existing novelty estimation methods plots for 30 randomly selected quotes, drawn from three categories: classical Chinese poetry, modern Chinese prose, and English.}
    \label{fig:token-level}
\end{figure*}

Both baselines are biased: the uniform average is dominated by long continuation segments, while the TopK average is overly sensitive to an arbitrary prefix cutoff and may miss later turning tokens entirely.
By contrast, our novelty-token method assigns weights based on turning points of the self-perplexity trajectory, simultaneously capturing salient changes and down-weighting flat continuation regions.
Under this weighting, the model instead recommends:
\begin{quote}
    \small
    \begin{CJK}{UTF8}{gbsn}       
        (中文) 但使书种多，会有岁稔时。
        
        (English) If only we sow many seeds of learning, a season of abundance will surely come.
    \end{CJK}
\end{quote}
This quotation is both contextually appropriate and genuinely novel, illustrating that our method offers a more robust and general treatment of auto-regressive continuation bias than uniform or TopK averaging.
\subsection{Is continuation bias a key factor behind baseline failures?}
\label{appendix:continuation-bias-analysis}
To better understand whether the continuation bias we identify is indeed a major factor underlying the failures of existing novelty-estimation methods, we conduct a controlled token-level analysis across 30 randomly sampled quotations. As shown in Figure~\ref{fig:token-level}, methods that directly rely on likelihood-based signals—such as Surprisal~\citep{surp} and KL-Divergence~\citep{kl}—exhibit a consistent pattern: once the model enters a locally predictable phrase, the remaining tokens receive artificially low novelty scores, even when the quotation is globally unexpected. This aligns with the findings of continuation bias, where auto-regressive language models tend to over-commit to familiar continuations, thereby distorting novelty estimates at the sequence level.

Interestingly, metrics that do not depend on auto-regressive probability, such as Self-BLEU~\citep{selfbleu} and Embedding-Distance~\citep{emb}, do not show such degradation, which further confirms that the observed issue stems from the probabilistic continuation mechanism rather than from the quotations themselves. It is worth noting that Self-BLEU and Embedding-Distance are not affected by auto-regressive continuation bias, yet they \textbf{still lag behind} our estimator in both novelty scores and downstream ranking metrics (Table~\ref{tab:ab3}). This is because they operationalize a different notion of ``novelty". Self-BLEU primarily measures \textbf{lexical diversity} with respect to reference quotations. Conversely, truly insightful quotations often reuse common vocabulary, leading Self-BLEU to underestimate their novelty. Embedding-Distance treats novelty as \textbf{global semantic distance} in an embedding space. In other words, these metrics capture unconditional dissimilarity rather than \textbf{context-conditioned surprise}, which is exactly what our logit-based novelty-token estimator is designed to model. 
This explains why they are less aligned with human preferences for ``unexpected yet rational'' quotations, despite not suffering from continuation bias.

\textbf{Importantly, our goal here is not to claim that continuation bias is the sole reason existing methods fail}. Instead, our analysis highlights that continuation bias constitutes a systematic and previously overlooked source of error that affects a broad class of likelihood-based novelty estimators. By identifying this mechanism, we provide a principled explanation for why these methods underperform in quotation-recommendation settings, and motivate the design of our token-level novelty-token estimator, which explicitly mitigates this bias. To demonstrate this, we also applied the novelty token design to Surprisal and KL-Divergence and observed the results, as shown in Table~\ref{tab:ab3}. The results were improved to some extent, but still weaker than our method.

\section{Definition of Other Novelty Estimation Method}
\label{appendix:definition-of-other-novelty-estimation-method}
To verify that the proposed logit-based novelty is not the only way to capture ``unexpected yet rational'' quotes, we also experimented with several alternative novelty metrics. These methods are evaluated in the main paper (Section~\ref{sec:novelty-estimation}). Below we describe their definitions and the motivation for using each metric.

\subsection{Surprisal-based Novelty}

For a candidate quote $q = (x_1,\ldots,x_T)$ and context $C$, 
we define the average token surprisal as~\citep{surp}:
$$
\mathrm{Surprisal}(q)=
\frac{1}{T}\sum_{t=1}^T - \log P(x_t \mid C, x_{<t})
$$
This measures how unpredictable a token is in its context.
Higher average surprisal indicates that the model finds
the quote harder to predict, which can be associated with novelty.


\subsection{KL-Divergence between Prior and Conditional Distributions}

For each token we compute two probability distributions:
$P_{\mathrm{prior}}(\cdot|x_{<t})$ without context
and $P_{\mathrm{cond}}(\cdot|C,x_{<t})$ with context.
The novelty score is then the average KL divergence~\citep{kl}:
$$
\mathrm{KL}(q)=\frac{1}{T}\sum_{t=1}^T
D_{\mathrm{KL}}\big(
P_{\mathrm{prior}} \parallel P_{\mathrm{cond}}
\big)
$$
A larger distributional shift means the context makes the tokens
less expected, capturing a stronger ``surprise'' effect.

\subsection{Embedding-based Distance}

Let $e(q)$ be the embedding of a quote and 
$\mathcal{N}_k(q)$ its $k$ nearest neighbors in the corpus.
The embedding-based novelty is~\citep{emb}:
$$
\mathrm{Dist}(q)=
\frac{1}{k}\sum_{q' \in \mathcal{N}_k(q)}
\big(1-\cos(e(q),e(q'))\big)
$$
Quotes farther away from known ones in semantic space
are considered more novel.

\subsection{Self-BLEU Diversity}

We also compute the BLEU score between a quote $q$ and its closest match $q^*$ in the training corpus~\citep{selfbleu}:
$$
\mathrm{Self\text{-}BLEU}(q) = 1 - \mathrm{BLEU}(q,q^*)
$$
A higher Self-BLEU score indicates lower lexical overlap,
thus higher diversity and novelty.

\subsection{Uniform/TopK Average}
In our method, we propose performing token-level weighting of the novelty tokens when computing the final novelty score. To further verify the effectiveness of this design, we compare two weighting schemes:

(1) Uniform Average: uniformly weight all tokens
$$
S_{uniform} = \frac{1}{T}\sum_{t=1}^T R_t.
$$

(2) TopK Average: only weight the top $K$ tokens (here we set $K=5$ tokens)
$$
S_{topk} = \frac{1}{K}\sum_{t=1}^K R_t.
$$



\section{Prompt}

\subsection{Prompt for LLM-as-Judge}
\label{appendix:prompt-llm-as-judge}
Overall, we evaluate the performance of recommending a quotation by asking a strong LLM to score it along two dimensions: contextual matching and novelty. We empirically verify that this LLM-as-judge setup is effective and aligns well with human judgments. Below we present the prompts used for rating mathcing (appropriateness) and novelty, respectively.

\vspace{1em}
 \noindent\textbf{Prompt 1.1: Semantic Matching Evaluation}

\noindent\rule{\linewidth}{0.4pt} 
\begin{quote}
    \small
    \textit{Task prompt}

        You are an expert evaluator. Given a ``context'' text and a single ``candidate quote,'' rate the quote on the dimension below:\\
        \\
        \textbf{Semantic Matching (1--5)}: How well does this quote align with the main topic, argument, or intent of the context?\\
        (1 = off-topic; 5 = directly and indispensably connected)\\

        \textit{Output requirements}

        Please output in this YAML format:\\
        \\
        matching:\\
          reason: brief justification for your matching score\\
          score: Y\\
        \\
        Note:\\
        - If the quote is in Chinese, write the reason in \textbf{Chinese}; otherwise, write it in \textbf{English}.\\
        - Only evaluate this single dimension.\\
        - Please first give the reason and then give the score.\\
        \\
        Example1:\\
        Context: ``In personal image matters, traditional Confucianism advocates achieving personal improvement through self-cultivation and moral perfection. Now, with the rapid development of the Internet and the rise of social media, people are increasingly concerned about how others perceive them.''\\
        Quote: ``Your brand is what people say about you behind your back.''\\
        Deep Meaning of Quote: ``Expresses that true reputation exists in spaces we cannot control, reflected in others' genuine evaluations behind our backs.''\\
        Output:\\
        matching:\\
          reason: ``The quote highly aligns with the context's argument about 'image being derived from others' perceptions in the social media age,' providing an appropriate and profound supplement.''\\
          score: 5\\
        \\
        Example2:\\
        Context: ``In times of uncertainty and crisis, leaders are expected to provide clarity, calm, and a sense of direction. Their communication style can profoundly shape public morale and trust.''\\
        Quote: ``A leader is one who knows the way, goes the way, and shows the way.''\\
        Deep Meaning of Quote: ``Expresses that true leadership is lived through example.''\\
        Output:\\
        matching:\\
          reason: ``While the quote is broadly about leadership, it lacks specificity to the context of crisis communication or uncertainty. It fits the topic loosely but doesn't enrich the argument.''\\
          score: 3\\

        \textit{Input}
        
            ---INPUT---\\
            Context: ``\textlangle context\textrangle''\\
            Quote: ``\textlangle quote\textrangle''\\
            Deep Meaning of Quote: ``\textlangle deepmeaning\textrangle''\\
            \\
            Please start your evaluation and provide the output in the specified YAML format without other information or strings.\\
            ---OUTPUT---\\
\end{quote}
\noindent\rule{\linewidth}{0.4pt} 

\vspace{1em}
 \noindent\textbf{Prompt 1.2: Novelty Evaluation}

\noindent\rule{\linewidth}{0.4pt} 
\begin{quote}
    \small
    \textit{Task prompt}
 \textit{Task prompt} 

    You are an expert evaluator. Given a ``context'' text and a single ``candidate quote,'' rate the quote on the dimension below:\\
    \\
    \textbf{Surprise Novelty (1--5)}: How surprising, clever, or ``wow-worthy'' is this quote in light of the context?\\
    (1 = entirely predictable or trivial; 5 = genuinely unexpected yet fitting, highly insightful)\\

    \textit{Output requirements}
    
    Please output in this YAML format:\\
    \\
    novelty:\\
      reason: brief justification for your novelty score\\
      score: X\\
    \\
    Note:\\
    - If the quote is in Chinese, write the reason in \textbf{Chinese}; otherwise, write it in \textbf{English}.\\
    - Only evaluate this single dimension.\\
    - Please firstly give the reason and then give the score.\\
    \\
    Example1:\\
    Context: ``In personal image matters, traditional Confucianism advocates achieving personal improvement through self-cultivation and moral perfection. Now, with the rapid development of the Internet and the rise of social media, people are increasingly concerned about how others perceive them.''\\
    Quote: ``Your brand is what people say about you behind your back.''\\
    Deep Meaning of Quote: ``Expresses that true reputation exists in spaces we cannot control, reflected in others' genuine evaluations behind our backs.''\\
    Output:\\
    novelty:\\
      reason: ``This quote reinterprets personal image through the modern 'brand' concept, offering a refreshing perspective while accurately capturing the impact of others' evaluations on self-perception in the social media age.''\\
      score: 5\\
    \\
    Example2:\\
    Context: ``In times of uncertainty and crisis, leaders are expected to provide clarity, calm, and a sense of direction. Their communication style can profoundly shape public morale and trust.''\\
    Quote: ``A leader is one who knows the way, goes the way, and shows the way.''\\
    Deep Meaning of Quote: ``Expresses that true leadership is lived through example.''\\
    Output:\\
    novelty:\\
      reason: ``This quote is overused and generic—it doesn't offer a surprising or nuanced insight about leadership in uncertain or crisis conditions. It's surface-level and predictable.''\\
      score: 2\\
      
        \textit{Input}
        
        ---INPUT---\\
        Context: ``\textlangle context\textrangle''\\
        Quote: ``\textlangle quote\textrangle''\\
        Deep Meaning of Quote: ``\textlangle deepmeaning\textrangle''\\
        Please start your evaluation and provide the output in the specified YAML format without other information or strings.\\
        ---OUTPUT---\\
\end{quote}
\noindent\rule{\linewidth}{0.4pt}

\subsection{Prompt for label agent}
\label{appendix:prompt-label-agent}
In the label agent (Section~\ref{sec:label-enhancement} and Appendix~\ref{appendix: label-agent}), we process it in 3 prompts (analysis and deep-meaning labeling, multi-round correction, and multi-dimensional label), as shown below.

\vspace{1em}
 \noindent\textbf{Prompt 2.1: Analysis and Deep-meaning Labeling}

\noindent\rule{\linewidth}{0.4pt} 
\begin{quote}
    \small
\textit{Task prompt (Analysis \& Deep Meaning)}

    Please act as an expert well-versed in English quotes. Perform a comprehensive and in-depth analysis of the following famous quote. Use the format below: \\
    \textlangle AA\textrangle ... \textlangle/AA\textrangle \\[2pt]
    Your analysis should include but is not limited to the following aspects: \\
    1. \textbf{Origin and Author Background} \\
    \quad Indicate who wrote this quote and briefly introduce the author's life and creative context. \\
    2. \textbf{Historical and Cultural Background} \\
    \quad Explain the historical era in which the quote was created and whether there were any specific cultural or societal contexts surrounding it. \\
    3. \textbf{Line-by-Line / Word-by-Word Interpretation} \\
    \quad Provide concise interpretations of each key image or word in the quote. \\
    4. \textbf{Emotional Connotation} \\
    \quad Analyze the underlying emotions in the quote, such as friendship, loneliness, melancholy, etc. \\[2pt]
    \textbf{Note:} Please analyze the quote based on the given context and any additional information, but there is no need to interpret the broader context itself. \\[4pt]

    \textit{Deep meaning}

    Based on the above analysis, extract the deeper meaning of the quote and summarize it in fewer than 50 characters. Focus on the abstract meaning, not the concrete object or scene. Use the format below: \\
    \textlangle DM\textrangle Expresses that … \textlangle/DM\textrangle \\[2pt]
    Example: \\
    \textlangle DM\textrangle Expresses that true learning and growth come from active engagement and firsthand experience.\textlangle/DM\textrangle \\
    \textlangle DM\textrangle Expresses that holding onto the past or dwelling on today's troubles is ultimately futile because time moves forward.\textlangle/DM\textrangle \\[4pt]

    \textit{Input}

    ---INPUT---\\
    Quote to Analyze: \\
    \{quote\} \\
    Author: \\
    \{author\} \\
    Additional Information: \\
    \{info\} \\[4pt]

    \textit{Output}

    Please provide your output in a clear structure, refined language, and well-organized layout: \\
    1.~Analysis Result: \textlangle AA\textrangle Text \textlangle/AA\textrangle \\
    2.~Deep Meaning: \textlangle DM\textrangle Text \textlangle/DM\textrangle \\
    Now generate: \\
\end{quote}
\noindent\rule{\linewidth}{0.4pt} 

\vspace{1em}
\noindent\textbf{Prompt 2.2: Multi-round correction}

\noindent\rule{\linewidth}{0.4pt} 

\begin{quote}
    \small
    Please apply multi-round self-correction to your answer:\\
    1. Check for superficial or shallow explanations.\\
    2. Check for over-interpretation or unsupported assumptions.\\
    3. Check for logical gaps or inconsistencies.\\
    
    If you think this instruction itself is incorrect or invalid, just answer "No". Otherwise, answer "Yes".\\

\end{quote}
\noindent\rule{\linewidth}{0.4pt} 

\vspace{1em}
\noindent\textbf{Prompt 2.3: Multi-dimensional label}

\noindent\rule{\linewidth}{0.4pt} 

\begin{quote}
    \small
    \textit{Task prompt (Label generation)} 

    Please act as an expert well-versed in English quotes. Based on the quotation and its deep-meaning explanation (if provided), assign fine-grained, multidimensional labels to support precise semantic search. Use the format below: \\
    \textlangle LB\textrangle JSON \textlangle/LB\textrangle \\[4pt]

    \textit{Labeling dimensions (all keys in English)} 

    1.~\textbf{core\_domains} (1--2 items) \\
    \quad Choose from predefined domains, e.g.\ [``Life Philosophy'', ``Knowledge \& Learning'', ``Success \& Achievement'', ``Love \& Family'', ``Separation \& Longing'', ``Spiritual Solace'', ``Politics \& War''], etc. \\
    \quad Example: [``Separation \& Longing''] \\[2pt]
    2.~\textbf{core\_insights} (1--3 items) \\
    \quad Capture the essential behavioral advice or insight conveyed by the quote as short verb phrases or core statements. Avoid vague nouns. \\
    \quad Example: [``Expressing emotion through letters'', ``Caring for others' well-being''] \\[2pt]
    3.~\textbf{applicability} (0--2 items) \\
    \quad The most relevant scenario(s) or audience(s) for applying this quote. \\
    \quad Example: [``Homesick traveler writing home''] \\[2pt]
    4.~\textbf{core\_values} (1--2 items) \\
    \quad The values or attitudes implied or advocated by the quote, refined within the selected core domain(s). \\
    \quad Example: [``Care'', ``Filial Piety''] \\[2pt]
    5.~\textbf{metaphors} (1 item) \\
    \quad Identify the most representative metaphor or symbol in the quote. \\
    \quad Example: [``Letter'', ``Friendship''] \\[2pt]
    6.~\textbf{style} \\
    \quad The primary rhetorical device or stylistic feature. \\
    \quad Example: [``Rhetorical Question''] \\[2pt]
    7.~\textbf{sentiment\_tone} (1--2 items) \\
    \quad The main emotional tone(s) or mood conveyed by the quote. \\
    \quad Example: [``Melancholy'', ``Longing''] \\[4pt]

    \textit{Input} 
    
    ---INPUT---\\
    Quote: \\
    \{quote\} \\
    Author: \\
    \{author\} \\
    Additional Information: \\
    \{info\} \\
    Deep Meaning (optional): \\
    \{deep\_meaning\} \\[4pt]

    \textit{Output} 

    Please output only a single JSON object wrapped in the following tag: \\
    \textlangle LB\textrangle \{ \\
    \quad "core\_domains": [...], \\
    \quad "core\_insights": [...], \\
    \quad "applicability": [...], \\
    \quad "core\_values": [...], \\
    \quad "metaphors": [...], \\
    \quad "style": "...", \\
    \quad "sentiment\_tone": [...] \\
    \} \textlangle/LB\textrangle \\
    Now generate: \\
\end{quote}
\noindent\rule{\linewidth}{0.4pt}

\subsection{Questionnaire}
\label{appendix:prompt-questionnaire}
Below we present the full questionnaire used in Appendix~\ref{appendix:user-study}.
The original survey was administered online; questions are shown here in English. And we will randomly select the following quotations from KB. \\
\noindent\rule{\linewidth}{0.4pt} 

\begin{quote}
    \small
\noindent\textbf{Welcome!}
\begin{quote}
Thank you for participating in this survey about how people use and think about quotations in writing. 

In this questionnaire, you will:
\begin{itemize}
    \item answer a few questions about your writing background,
    \item rate several example quotations in context,
    \item tell us how you would prefer to use quotations in different writing scenarios, and
    \item optionally share your own views about what makes an ``ideal'' quotation.
\end{itemize}

There are no right or wrong answers. We are only interested in your honest opinions and preferences.

The survey takes about 10--15 minutes to complete.
Your responses will be used for research purposes only and will be analyzed in anonymized form.

By clicking ``Next'' and starting the survey, you confirm that:
\begin{itemize}
    \item you are at least 18 years old, and
    \item you consent to participate in this anonymous study.
\end{itemize}
\end{quote}

\subsubsection*{Part A: Demographics and Writing Background}

\noindent\textbf{Q1. Age group / Work field}

Which age group are you in?

\begin{itemize}
    \item 18--24
    \item 25--34
    \item 35--44
    \item 45--54
    \item 55+
\end{itemize}

What is your primary work field?

\begin{itemize}
    \item Education
    \item Research
    \item Industry
    \item Other: \underline{\hspace{3cm}}
\end{itemize}

\vspace{0.5em}
\noindent\textbf{Q2. Primary language for writing}

Which language do you mainly use when you write longer texts (e.g., essays, reports, blog posts)?

\begin{itemize}
    \item Chinese
    \item English
    \item Both Chinese and English
    \item Other: \underline{\hspace{3cm}}
\end{itemize}

\vspace{0.5em}
\noindent\textbf{Q3. Writing frequency}

How often do you write long-form texts (e.g., essays, reports, blog posts, stories)?

\begin{itemize}
    \item Almost never
    \item A few times a year
    \item About once a month
    \item About once a week
    \item Several times a week or more
\end{itemize}

\vspace{0.5em}
\noindent\textbf{Q4. Typical writing domains} (multiple choice)

In which domains do you write most often?

\begin{itemize}
    \item School essays / assignments
    \item Academic papers / theses
    \item Blogs or long social media posts
    \item Business reports or presentations
    \item Internal company emails / announcements
    \item Legal or policy documents
    \item Medical or health-related documents
    \item Creative writing (fiction, poetry, scripts, etc.)
    \item Other: \underline{\hspace{3cm}}
\end{itemize}

\vspace{0.5em}
\noindent\textbf{Q5. Familiarity with using quotations}

When you write, how familiar are you with using quotations (e.g., famous sayings, lines from books or movies)?

(1 = I rarely use quotations, 5 = I frequently use them and think carefully about which ones to choose.)

\begin{itemize}
    \item 1 -- I rarely use quotations
    \item 2
    \item 3
    \item 4
    \item 5 -- I very often use quotations and think a lot about them
\end{itemize}

\subsubsection*{Part B: Views on ``Appropriateness'' and ``Novelty''}

\noindent\textit{Explanation shown to participants:}

\begin{quote}
\small
In this survey we talk about two aspects of a quotation:

\textbf{Appropriateness} (or ``fit''):  
How well the quotation matches the surrounding text and context, in terms of meaning and logic.
A highly appropriate quotation feels natural and makes sense where it appears.

\textbf{Novelty} (or ``unexpectedness''):  
To what extent the quotation feels fresh, not clich\'ed, and somewhat surprising or eye-opening in this context, without becoming nonsense.

In the questions below, please think about these two aspects separately.
\end{quote}

\vspace{0.5em}
\noindent\textbf{Q6. Importance of appropriateness}

In your opinion, how important is \emph{contextual appropriateness} for an ``ideal'' quotation?

(0 = not important at all, 10 = absolutely essential)

\begin{quote}
Please choose a number from 0 to 10: \underline{\hspace{1cm}}
\end{quote}

\vspace{0.5em}
\noindent\textbf{Q7. Importance of novelty}

In your opinion, how important is \emph{novelty / unexpectedness} for an ``ideal'' quotation?

(0 = not important at all, 10 = extremely important)

\begin{quote}
Please choose a number from 0 to 10: \underline{\hspace{1cm}}
\end{quote}

\vspace{0.5em}
\noindent\textbf{Q8. Relationship between appropriateness and novelty}

For each statement below, please indicate how much you agree or disagree.

(1 = strongly disagree, 5 = strongly agree)

\begin{itemize}
    \item[(a)] ``A good quotation must be appropriate for the context first; if it is not appropriate, it cannot be good.''
    \item[(b)] ``Once a quotation is appropriate, I tend to like it more if it feels less clich\'ed and a bit more original.''
    \item[(c)] ``Appropriateness and novelty feel like two dimensions to me: one makes the quotation `make sense', the other makes it `stand out'.''
    \item[(d)] ``As long as a quotation is novel enough, it does not really matter whether it fits the context.''
    \item[(e)] ``As long as a quotation fits the context, I do not care whether it is clich\'ed or original.''
\end{itemize}

(For each item: 1 = strongly disagree, 2 = disagree, 3 = neutral, 4 = agree, 5 = strongly agree.)

\vspace{0.5em}
\noindent\textbf{Q9. Intuitive picture of the two dimensions}

Please imagine a two-dimensional plane:

\begin{itemize}
    \item Horizontal axis: Appropriateness (from not appropriate to very appropriate)
    \item Vertical axis: Novelty (from very clich\'ed to very surprising)
\end{itemize}

Which statement best matches your view of an ``ideal'' quotation? (choose one)

\begin{itemize}
    \item[A.] Ideally, a quotation is in the \textbf{top-right corner}: both appropriate and at least somewhat novel.
    \item[B.] Ideally, a quotation is in the \textbf{bottom-right corner}: as long as it is very appropriate, novelty does not matter.
    \item[C.] Ideally, a quotation is in the \textbf{top-left corner}: as long as it is very novel, I can tolerate it being a bit forced.
    \item[D.] None of the above (please briefly explain): \underline{\hspace{4cm}}
\end{itemize}

\subsubsection*{Part C: Direct Comparison Questions}

\noindent\textbf{Q10. When appropriateness is the same}

Suppose you have two quotations that are \emph{equally appropriate} for your text (both fit the context very well):

\begin{itemize}
    \item Quote A: very common and widely used; most readers have seen it many times.
    \item Quote B: less common and somewhat more original, but still fully appropriate and reasonable.
\end{itemize}

In this situation, which quotation would you \emph{tend to choose}?

(1 = definitely choose A, 5 = definitely choose B)

\begin{itemize}
    \item 1 -- I would definitely choose the more common A
    \item 2 -- I would usually choose A
    \item 3 -- It depends / no clear tendency
    \item 4 -- I would usually choose the more original B
    \item 5 -- I would definitely choose the more original B
\end{itemize}

\noindent\textbf{Q10-Reason (optional free text).}  
Why would you make this choice?

\begin{quote}
\underline{\hspace{4cm}}
\end{quote}

\vspace{0.5em}
\noindent\textbf{Q11. When you must trade off appropriateness and novelty}

Now consider a slightly extreme situation. You have two options:

\begin{itemize}
    \item Quote C: very appropriate and fully makes sense in context, but slightly plain or clich\'ed.
    \item Quote D: very novel and rarely seen, but a bit stretched for the context (not completely wrong, but somewhat indirect or ``forced'').
\end{itemize}

If you had to choose \emph{one} quotation to include in your writing, what would you tend to choose?

(1 = definitely choose C, 5 = definitely choose D)

\begin{itemize}
    \item 1 -- Definitely choose the more appropriate C, even if it is boring
    \item 2 -- More likely C
    \item 3 -- It depends / not sure
    \item 4 -- More likely the more novel D, even if slightly forced
    \item 5 -- Definitely choose the more novel D
\end{itemize}

\noindent\textbf{Q11-Reason (optional free text).}  
Please briefly explain your reasoning:

\begin{quote}
\underline{\hspace{4cm}}
\end{quote}

\vspace{0.5em}
\noindent\textbf{Q12. Ranking different types when all are appropriate}

Suppose you have three types of quotations, all of which you consider \emph{appropriate} (e.g., you would rate their appropriateness at 4 or 5 out of 5):

\begin{itemize}
    \item Quote E: very common, very safe, but somewhat ordinary.
    \item Quote F: somewhat original, with a slightly different way of expressing the idea.
    \item Quote G: more clearly original, giving a stronger feeling of ``freshness'', but still understandable and on-topic.
\end{itemize}

In your actual writing, how would you usually \emph{rank} these three types of quotations by preference (from most preferred to least preferred)?

\begin{quote}
My typical order would be: \underline{\hspace{5cm}} (for example: F $>$ G $>$ E)
\end{quote}

\vspace{0.5em}
\noindent\textbf{Q13. Which statement is closest to your true preference?}

Please choose the one that best describes you:

\begin{itemize}
    \item[a.] As long as a quotation feels appropriate, I do not care much whether it is common or original.
    \item[b.] Once a quotation is appropriate, I still tend to prefer those that feel a bit less clich\'ed and more original.
    \item[c.] I actively hope quotations will give readers some sense of surprise, as long as they are not wildly off-topic.
    \item[d.] None of the above (please briefly explain): \underline{\hspace{4cm}}
\end{itemize}

\subsubsection*{Part D: Preferences Across Writing Scenarios}

\noindent\textbf{Q14. Preference for novelty in different writing scenarios}

For each type of writing below, imagine that you already have several quotations that are all \emph{appropriate} for your text.
Some are more common and ``safe'', others are more novel.

Please indicate which kind of quotation you would normally prefer in this scenario:

(1 = strongly prefer common and safe quotations, 5 = strongly prefer novel quotations)

\begin{enumerate}[label=(\alph*)]
    \item Creative writing (short stories, fiction)
    \item Personal essays or reflections (about your own experiences, feelings, or growth)
    \item Opinion pieces / commentary (on news, social issues, trends)
    \item Book / movie / music reviews
    \item Ordinary school essays / exam essays
    \item Academic research papers
    \item Business reports or presentations
    \item Internal company emails / announcements
    \item Legal contracts / policy documents
    \item Medical or health information leaflets
\end{enumerate}

For each scenario, participants select one value from 1 to 5:
1 = strongly prefer common and safe quotations, 5 = strongly prefer novel, unexpected yet rational quotations.

\subsubsection*{Part E: Self-reported Behavior and Open-ended Feedback}

\noindent\textbf{Q15. Avoiding clich\'es}

When you write, do you consciously avoid quotations that feel too clich\'ed or ``cheesy''?

\begin{itemize}
    \item Almost never; I am fine with very classic quotations.
    \item Sometimes.
    \item I often try to avoid very clich\'ed quotations.
    \item I almost always avoid very clich\'ed quotations.
\end{itemize}

\vspace{0.5em}
\noindent\textbf{Q16. Removing quotations because they feel too ordinary}

Have you ever \emph{removed} a quotation from your draft simply because it felt too ordinary or overused?

\begin{itemize}
    \item Never
    \item Once or twice
    \item Several times
    \item Very often
\end{itemize}

\vspace{0.5em}
\noindent\textbf{Q17. Open-ended: What makes an ``ideal'' quotation?}

In your own words, what makes a quotation feel ``ideal'' or ``memorable'' in a piece of writing?

\begin{quote}
\underline{\hspace{4cm}}
\end{quote}

\vspace{0.5em}
\noindent\textbf{Q18. Open-ended: Is being unexpected important?}

Do you think being unexpected (novel) is important for quotations? Why or why not?

\begin{quote}
\underline{\hspace{4cm}}
\end{quote}

\end{quote}
\noindent\rule{\linewidth}{0.4pt}

\end{document}